\documentclass[useAMS,usenatbib,a4paper]{mn2e}
\usepackage{times}
\title[Giant Molecular Cloud interactions]
{
The frequency and nature of `cloud-cloud collisions' in galaxies
}
\author[Dobbs]
{C. L. Dobbs\thanks{E-mail:
dobbs@astro.ex.ac.uk}$^{1}$, J. E. Pringle$^2$, and A. Duarte-Cabral$^1$ \\
$^1$ School of Physics and Astronomy, University of Exeter, Stocker Road, Exeter, EX4 4QL, UK \\
$^2$ Institute of Astronomy, Madingley Road, Cambridge, CB3 0HA}
\usepackage{amssymb}
\usepackage{bm}
\usepackage[pdftex]{graphicx}
\usepackage{epsfig}
\usepackage{multirow}

\begin{document}
\label{firstpage}
\date{\today}

\pagerange{\pageref{firstpage}--\pageref{lastpage}} \pubyear{2014}

\maketitle

\begin{abstract}
We investigate cloud-cloud collisions, and GMC evolution, in hydrodynamic simulations of isolated galaxies. The simulations include heating and cooling of the ISM, self--gravity and stellar feedback. Over timescales $<5$ Myr most clouds undergo no change, and mergers and splits are found to be typically two body processes, but evolution over longer timescales is more complex and involves a greater fraction of intercloud material.
We find that mergers, or collisions, occur every 8-10 Myr (1/15th of an orbit) in a simulation with spiral arms, and once every 28 Myr (1/5th of an orbit) with no imposed spiral arms. Both figures are higher than expected from analytic estimates, as clouds are not uniformly distributed in the galaxy. Thus clouds can be expected to undergo between zero and a few collisions over their lifetime. We present specific examples of cloud--cloud interactions in our results, including synthetic CO maps. We would expect cloud--cloud interactions to be observable, but find they appear to have little or no impact on the ISM. Due to a combination of the clouds' typical geometries, and moderate velocity dispersions, cloud--cloud interactions often better resemble a smaller cloud nudging a larger cloud. Our findings are consistent with the view that spiral arms make little difference to overall star formation rates in galaxies, and we see no evidence that collisions likely produce massive clusters. However, to confirm the outcome of such massive cloud collisions we ideally need higher resolution simulations.
\end{abstract}


\section{Introduction}
Cloud-cloud collisions have long been thought to play an important role in both the growth of molecular clouds, and potentially the star formation rate in galaxies. However the nature of cloud-cloud collisions, how we identify them, and whether they are sufficiently frequent to contribute to cloud growth, are still unanswered questions. Establishing examples of cloud-cloud collisions observationally is difficult, and even those studies which claim to find cloud-cloud collisions are not definitive. 

Many early papers proposed cloud-cloud collisions as a means of building up more massive Giant Molecular Clouds (GMCs) from smaller molecular clouds \citep{Field1965,Scoville1979,Norman1980,Tomisaka1984,Tomisaka1986,Kwan1983,Kwan1987,Roberts1987}. One criticism of this earlier work was that the expected timescale between cloud-cloud collisions was very long, of order 100 Myr or more \citep{Blitz1980}. It was subsequently proposed \citep{Casoli1982,Kwan1983,Dobbs2008} that cloud-cloud collisions would be enhanced in spiral arms. Calculations in the 1980's modelling clouds orbiting a galaxy found an enhancement of a factor of a few in the spiral arms \citep{Kwan1983,Tomisaka1984} though they include neither hydrodynamics or self--gravity. \citet{Dobbs2008} carried out hydrodynamic calculations, but did not quantitatively investigate the effect of spiral arms on cloud-cloud collisions. \citet{Tasker2009} computed the frequency in hydrodynamic calculations of galaxies with no spiral potential, obtaining a frequency of 1 collision per 1/4 orbit. They do not discuss in depth why this differs by up to an order of magnitude from the theoretical work, but do argue that self--gravity is important in their calculations. With an imposed spiral and bar potential, \citet{Fujimoto2014} find merger rates as high as one every 2 or 3 Myr (1 collision per 1/40th of an orbit).

Cloud-cloud collisions have also been supposed to be important for star formation. The Schmidt Kennicutt relation can be expressed in a form which includes the angular velocity, $\Omega$ \citep{Kennicutt1998, Wyse1986, Wyse1989, Silk1997, Tan2000}. In this instance, if the star formation rate is assumed to arise from cloud-cloud collisions, then these will be proportional to the shear of the disk, leading to a Schmidt relation of the form $\Sigma_{SFR} \propto \epsilon \Omega \Sigma$ where $\epsilon$ is the star formation efficiency.  Recently \citet{Inoue2013} have also suggested that cloud-cloud collisions lead in particular to massive star formation (see also discussion in \citealt{Longmore2014}). A number of observations of massive star clusters also display evidence that they  have formed as the result of cloud-cloud collisions \citep{Stolte2008,Furukawa2009,Torii2011,Fukui2014}.

In addition to the massive clusters above, a number of nearby smaller molecular clouds are thought to be in the process of colliding \citep{Higuchi2010,Galvan2010,Duarte2011,Nakamura2012}. The colliding clouds are primarily identified by blue and redshifted velocity fields. There is still some uncertainty as to whether these truly are cloud-cloud collisions, as the velocity fields are often quite complex, and may also reflect internal motions within the clouds. Nevertheless, we would statistically expect to see at least some cloud-cloud collisions in the galaxy, so some of these examples may well be truly colliding. \citet{Colombo2014} also consider the properties of GMCs in arm and inter-arm regions in M51, and suggest that differences in the mass function may reflect the occurrence of cloud-cloud collisions in the spiral arms.

As well as global numerical, or analytic studies of cloud-cloud collisions, there have been calculations of individual cloud-cloud collisions. These calculations have different predictions for the outcome of collisions. In some cases, cloud-cloud collisions can be quite violent, and largely destroy the natal clouds. This tends to be the case if the clouds collide with large Mach numbers, or velocities of at least several km s$^{-1}$ \citep{Hausman1981,Lattanzio1985,McLeod2011}. Although some previous work has considered the nature of collisions of different impact parameter \citep{Taff1973,Lattanzio1985}, the frequency of different impact parameters has not been considered in a global context. It is also not clear from the simulations so far what is the impact of collisions on the star formation rate, i.e. whether star formation increases (or even decreases) compared to isolated clouds, but the above studies indicate it may depend highly on the nature of the collision.
 
Whilst we have so far discussed cloud-cloud collisions, this picture does not reflect that the Interstellar Medium (ISM) is a continuous medium. We can only hypothesise cloud-cloud collisions (and indeed clouds themselves) by introducing some density cut off to define cloud boundaries. In reality, surrounding regions of the ISM will be interacting at lower densities than the cloud boundary. In such a scenario, the `collision' represents a converging flow, with likely increasing densities and decreasing sound speeds, until the cloud boundary is reached. The literature perhaps has also reflected the uncertainty of cloud cloud interactions. There are various terms to denote the interactions of clouds -- collisions, coalescence, mergers, agglomeration. Collisions are often used for all types of interaction, whereas coalescing clouds and mergers assume that the two clouds definitively join together. The term collision can often imply quite a violent interaction, though we note that this is probably not the case for molecular clouds. We further note that collisions can have any impact parameter. The term `agglomeration' was used in \citet{Dobbs2008}, and one or two other works, which is perhaps more indicative of random shaped objects sticking together than a collision. 

In this paper, we consider the nature of cloud evolution, in particular focusing on cloud mergers and their frequency. Compared to previous work, we provide a much more rigorous framework to identify cloud-cloud collisions, which is necessary when clouds evolve in time and space.
In \citet{Dobbs2013}, we highlighted the complexity of following the evolution of clouds in simulations, the frequent interactions, clouds splitting apart, or merging together as they evolve. These issues were in addition to the basic problem of how to define a cloud. Here we define interactions in terms of the transfer of mass between different time frames, and thus typically refer to collisions as mergers, as they involve mass transfer from two clouds into one.
In Section~2.2 we define 5 categories of cloud evolution. In Section~3 we show how clouds are divided into these categories, and over what timescales. In Sections 3.2 and 3.5 we determine cloud--cloud collision rates with and without strong spiral structure respectively, and compare both these measures to theory in Section~5. In Section~4 we investigate collisions/merger of massive clouds, and show specific examples from our calculations. 
In Section~6 we examine how one of our cloud mergers appears in H$_2$ and CO.  

\section{Details of simulations and Method}
\begin{figure}
\centerline{\includegraphics[scale=0.25, bb=380 80 700 580]{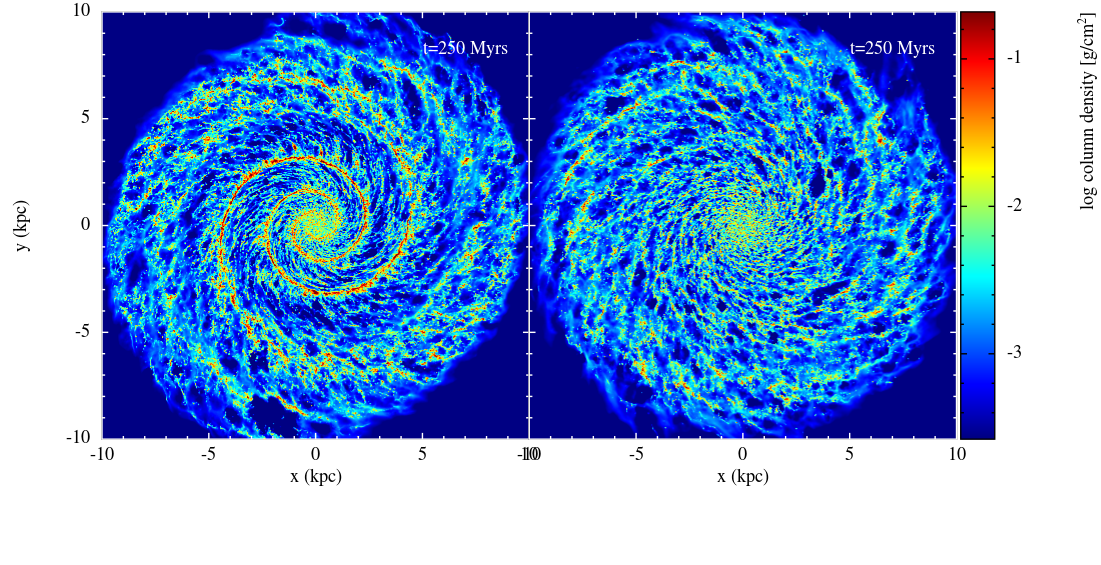}}
\caption{Column density plots for two simulations used in this paper are shown above, at a time of 250 Myr. The calculation shown in the left hand panel includes an $N=2$ imposed spiral perturbation, whereas for the calculation shown in the right-hand panel, gas is only subject to a symmetric logarithmic potential.}
\label{fig:columndensity}
\end{figure}

The simulations used in this paper are presented in \citet{Dobbs2013} and \citet{Dobbs2014}. These previous papers, and \citet{Dobbs2011new} show that the simulations reproduce well the large scale properties of the ISM (e.g. amounts of gas in cold and warm phases, scale heights), and properties of GMCs, compared to observations. The simulations are computed using the Smoothed Particle Hydrodynamics code sphNG \citep{Benz1990,Bate1995,PM2007}. We show results for two simulations, with and without a spiral potential. Otherwise the simulations have the same parameters. The surface density in each case is 8 M$_{\odot}$ pc$^{-2}$, and the number of particles 8 million, giving a mass per particle of 312.5 M$_{\odot}$. The calculations include self--gravity, cooling and heating (from \citealt{Glover2007}), and stellar feedback (from \citealt{Dobbs2011new}) with a star formation efficiency of 0.05. Star particles are not included in the simulations, there is only gas. The calculations include H$_2$ \citep{Dobbs2008} and CO \citep{Pettitt2014} formation. Both calculations include a logarithmic potential to provide a flat rotation curve; one calculation also includes the spiral potential of \citet{Cox2002}. Most of the results we show are for the case with the spiral potential. The spiral is imposed from $t=0$ in the simulations, and our analysis is carried out at a time of 250 Myr. The two calculations, with and without a spiral potential, are shown in Figure~\ref{fig:columndensity}.

Whilst we predominantly show results for the simulations above, we briefly mention a simulation with a higher surface density of 16 M$_{\odot}$ pc$^{-2}$, to appear in Duarte-Cabral et al. 2014 submitted., in Section~4. This otherwise has the same feedback, cooling and heating as the other simulations. This calculation used only 4 million particles. 

Further resolution studies, and tests of the feedback implementation in these models are shown in Dobbs 2014, submitted, where we model a section of a disc at much higher resolution. In Dobbs 2014, we are able to resolve clouds with 80 times more particles than shown here. These high resolution simulations verify the cloud properties (and properties of the ISM) found in the global models, and do not show a strong dependence on the details of stellar feedback. However because of the smaller coverage area, they contain too small a number of GMCs for the type of analysis we present here.

\subsection{Cloud selection}
\begin{figure}
\centerline{\includegraphics[scale=0.52, bb=60 0 500 670]{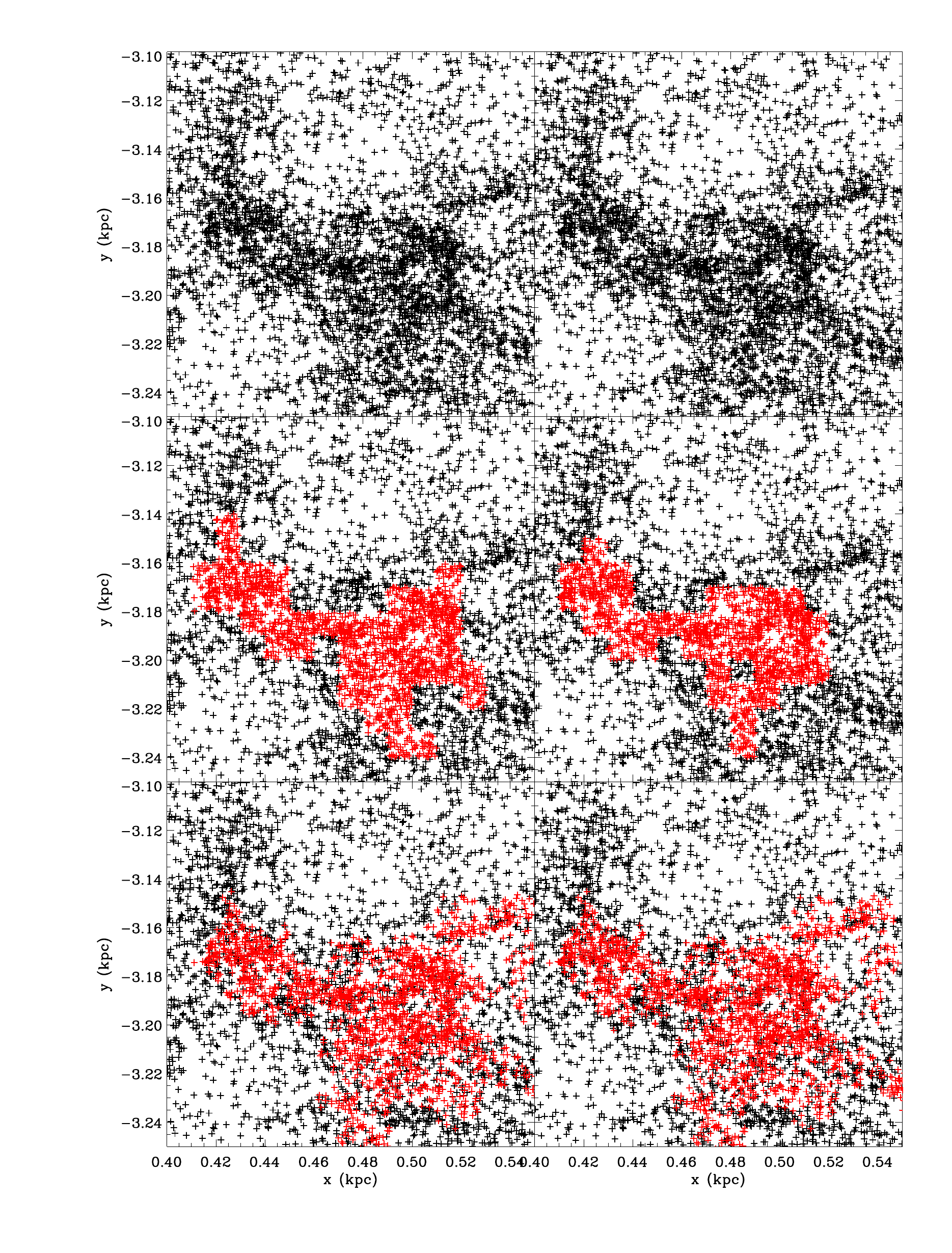}}
\caption{The particles are plotted for a small section of the galaxy (left) and again at a time of 0.1 Myr later (right). The middle panel shows a cloud as selected by the `grid--based' clump--finding algorithm. There are evident changes in the structure of the cloud even over a 0.1 Myr time frame, and the net change in mass is 10 \%. The bottom panels show clouds found using the `density-based' clump--finding algorithm (using $\rho_{min}=50$ cm$^{-3}$ and $l$=10 pc). The cloud is indistinguishable between the two time frames, as would be expected over such a short time period, and the net change in mass is 0.1 \%. The `density--based' algorithm is 3D, so some particles which may look like they should be part of the cloud in 2D may be further above or below the cloud particles in 3D.}
\label{fig:particleplots}
\end{figure}

In previous work \citep{Dobbs2014,Dobbs2008} we used a grid based approach to identify clouds, whereby we selected grid cells above a given surface density and group together all adjacent cells as a `cloud'. However whilst this grid based approach is sufficient for identifying clouds at a single snapshot (see Dobbs 2014, in prep.), we found it proved less suitable for studying clouds over time. This is because of the error associated with identifying cloud boundaries with a grid based approach. We show an illustration of this problem in Figure~\ref{fig:particleplots}, where we plot two clouds, identified 0.1 Myr apart. By looking at the particle distribution (top panels), we find no significant changes over 0.1 Myr. However the clump--finding algorithm picks out noticeably different structures (middle panels). When this algorithm is used, a cloud appears to change on shorter timescales than supposed by the actual particle distribution. We checked whether or not selecting clouds in the rotating frame of the potential contributed to this problem, but this was not found to have a big impact, rather the error lies in the conversion to a grid.

Instead here we adopt a `friends of friends' algorithm, which is non-grid based. We first select particles over a given (volume) density, $\rho_{min}$. We then group together all particles within a given length scale ($l$). This naturally produces clouds in 3D. There is some degeneracy between the density criterion $\rho_{min}$, and $l$, i.e. increasing $\rho_{min}$ gives very similar results to decreasing $l$. We again show the evolution of a cloud over 0.1 Myr in Figure~\ref{fig:particleplots} (lower panels). Unlike with the grid-based approach, there is negligible change in the structure of the cloud over such a small timescale, as would be expected. We use this method for the rest of the paper. 

\begin{table}
\centering
\begin{tabular}{c|cc|c|c|c|c}
 \hline 
$\rho_{min}$ (cm$^{-3}$) & $l$ (pc) & $\Sigma$ (M$_{\odot}$pc$^{-2}$) & $\%$ of gas in clouds \\
 \hline
50 & 10 & 45$\pm$10 & 3.3 \\
10 & 15 & 13$\pm$3 & 17 \\
\\
$\rho(H_2)_{min}$ (cm$^{-3}$) & $l$ (pc) & $\Sigma$ (M$_{\odot}$pc$^{-2}$) & $\%$ of gas in clouds \\
 \hline
10 & 15 & 45$\pm$10 & 2.8 \\
\hline
\end{tabular}
\caption{Table of the parameters associated with cloud selection. $\rho_{min}$ and $l$ are parameters used to identify clouds (see text). Then $\Sigma$ shows the average surface density of the clouds identified, and the standard deviation. The fraction of the total gas which lies in clouds is also shown. The lower part of the table shows parameters used when taking the molecular density rather than the total density, see Section~5}.
\label{tab:runs}
\end{table}

In Table~1 we show different parameters used to find clouds in the simulation. We select two separate populations of clouds, with different densities and surface densities. For our fiducial results, we take  $\rho_{min}=50$cm$^{-3}$ and $l=10$ pc (where $\rho$ reflects the total density). This gives clouds with a range of surface densities a little low compared to typical surveys, but similar to the Galactic $^{12}$CO clouds observed by \citet{Heyer2009}. The fraction of the total gas which lies in clouds is very small with these criteria, hence we also investigated clouds found using $\rho_{min}=10$cm$^{-3}$ and $l=15$ pc. As this yields clouds of unrealistically low surface densities compared to molecular clouds, the second criterion is mostly simply a comparison. However this could correspond to collisions of HI clouds, for example in the colliding flow scenario (e.g. \citealt{Vaz2007}).
Having a higher fraction of gas in clouds would also better reflect regions such as the inner parts of the Galaxy. For both cases we only consider clouds with masses over 1.5 $\times 10^4$ M$_{\odot}$, or over 50 particles. We repeated our analysis with higher mass limits, which also serves as a check for any dependence on resolution, but found similar results (see Section~4). Though obviously for calculating merger frequencies for massive clouds compared to the total number of clouds (see again Section~4), taking different lower limits for the massive clouds will yield different answers.

\subsection{How can clouds evolve?}
For simplicity we assume that interactions, evolution or fragmentation of clouds are typically at most a two body process. We can check the validity of this assumption later. But with this in mind, we can divide the possible evolution of clouds into a number of categories; `No change', `Merge', `Create', `Destroy' and `Split'. These represent all the possible outcomes between two different time frames. We determine the evolution of the clouds, and the numbers of clouds in each of these categories, by studying populations of clouds at two time frames, $T_0$ and $T_1$, where $T_1=T_0+\Delta T$. In all the cases here we take $T_0=250$ Myr. We can either study evolution forwards in time, or backwards in time, each giving different information about cloud formation, interactions or destruction. For example, mergers are obtained from the backwards evolution of the clouds, whereas splits are obtained from the forwards evolution.

\begin{figure}
\centerline{\includegraphics[scale=0.49,  bb=50 170 400 720]{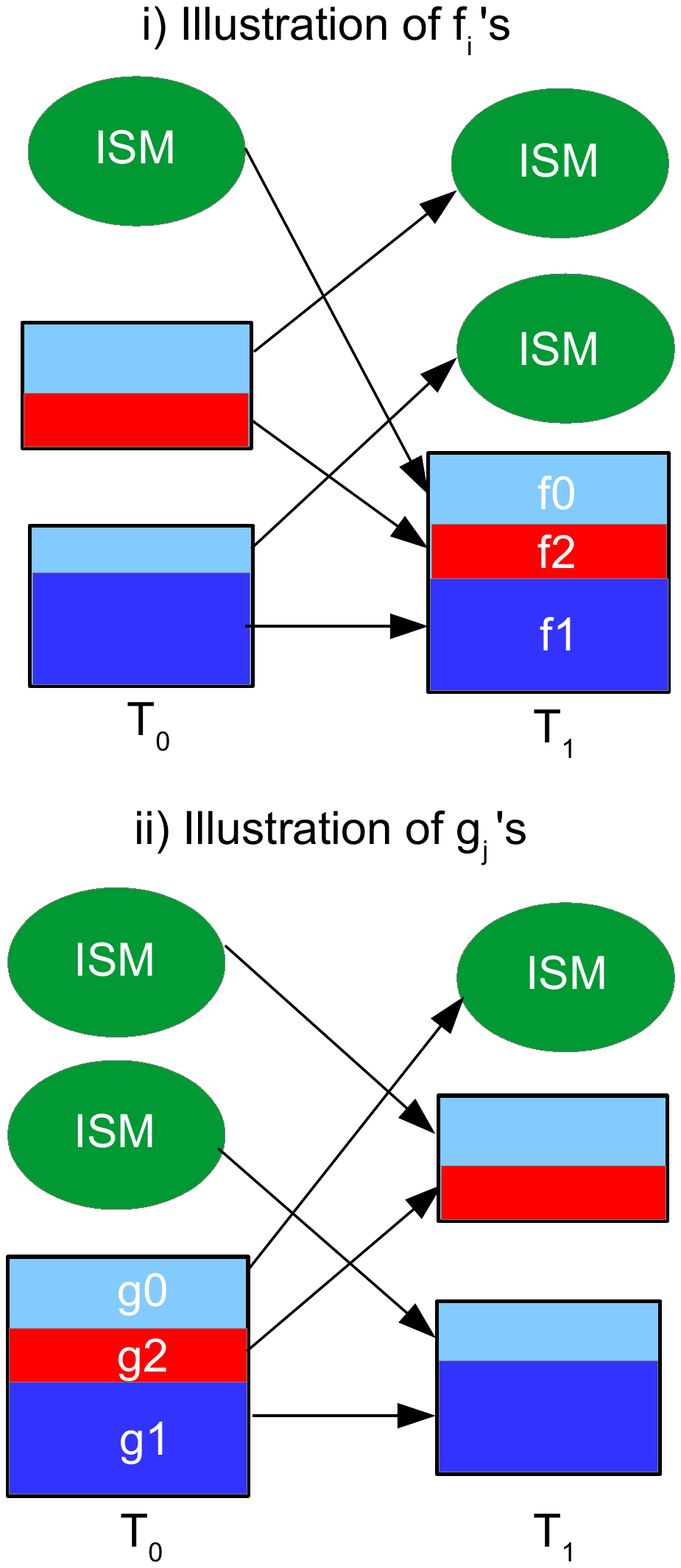}} 
\caption{This cartoon illustrates our definitions of $\bm{f}$ and $\bm{g}$. Clouds are represented by rectangles, which are subdivided into $f_i$ and $g_j$, and the ellipses represent intercloud ISM. The top panel represents backwards evolution. This particular example shows a merger. The $f_i$ indicate how gas in the cloud at $T_1$ was distributed in clouds and intercloud medium at time $T_0$.  The lower panel shows forwards evolution, and represents a split. Here the $g_j$ indicate how the gas in a cloud at $T_0$ is distributed in clouds and intercloud medium at $T_1$.}
\label{fig:figj}
\end{figure}

To determine how a population of clouds \textit{has} evolved, we consider the origin of clouds selected at time $T_1$. For each cloud at time $T_1$, we define $f_i$ as the fraction of that cloud which was in some cloud $i$ defined at time $T_0$ (see Figure~\ref{fig:figj}). We rearrange the $f_i$ such that they are decreasing, i.e. $f_1\geq f_2\geq f_3\ldots$ We also define $f_0=1-\Sigma^{\infty}_{i=1} f_i$, the amount of gas which was not in any cloud at time $T_0$. We can secondly consider how a population of clouds \textit{will} evolve, in this case from the time $T_0$. For this we define $g_j$ as the fraction of  a cloud existing at time $T_0$ which ends up in a cloud $j$ at time $T_1$ (see Figure~\ref{fig:figj}). We again rearrange $\bm{g}$ in order of decreasing $g_j$, and similarly define $g_0=1-\Sigma^{\infty}_{j=1} g_j$, the fraction of gas converted back into the intercloud ISM.

Using these definitions of  $\bm{f}$ and $\bm{g}$ we can then define the following categories of cloud evolution:
\begin{eqnarray*}
\textrm{Merge:} & f_1,f_2>0\\
\textrm{Create:} & f_i=0 \enspace \forall \thinspace i>0 \quad \textrm{and} \quad f_0>0\\
\textrm{Destroy:} & g_j=0 \enspace \forall \thinspace j>0 \quad \textrm{and} \quad g_0>0\\
\textrm{Split:} & g_1,g_2>0
\end{eqnarray*}
The `Create' category describes clouds that do not exist at time $T_0$ which appear by time $T_1$, whilst conversely `Destroy' describes clouds which exist at $T_0$, but whose material has returned to the intercloud medium by $T_1$.  If we are assuming that cloud evolution involves no more than two bodies, then we consider a `Merge' to occur when two clouds present at $T_0$ have merged into a single cloud at $T_1$. Similarly a split occurs when one cloud present at $T_0$ has split into two clouds at $T_1$. We define the number of clouds created as $N_C$, the number destroyed as $N_D$, and the number of mergers and splits as $N_M$ and $N_S$ respectively. Note that $N(T_1)-N(T_0)=N_C+N_S-N_D-N_M$ and that this gives a check on the two body assumption.

We have not yet assigned a `No change' category. This is complicated because this category requires knowing about the forward and backward evolution simultaneously, whereas we only know the $g_j$ at time $T_0$ and $f_i$ at time $T_1$. At this point it is helpful to introduce two auxiliary categories:
\begin{eqnarray*}
\textrm{No change} + \textrm{Split:} & f_i=0 \enspace \forall \thinspace i>1 \quad \textrm{and} \quad f_1>0 \\
\textrm{No change} + \textrm{Merge:} & g_j=0 \enspace \forall \thinspace j>1 \quad \textrm{and} \quad g_1>0 
\end{eqnarray*}
which cover unions of the categories mentioned earlier.
Note that categories Create, Merge and No change$+$Split cover all possible origins of clouds at $T_1$, whilst Destroy, Split and No change$+$Merge cover all possible evolution of clouds from $T_0$. With these auxiliary categories, and our initial assumption that splits or merges only involve two clouds, we can then determine the number of clouds in the `No change' category as:
\begin{eqnarray*}
\textrm{No change} \enspace N_{NC}:  & =N_{NC+S}-2N_S  \quad  or \quad  N_{NC+M}-2N_M.
\end{eqnarray*}
Again we have a check on the two body assumption.

In practice we identify clouds in these various categories by searching through the clouds at $T_0$ and $T_1$ to find those which contain the same particles. For example, if multiple clouds at $T_0$ contain the same particles as a cloud at $T_1$ we identify this as a merger. Similarly multiple clouds at $T_1$ that contain particles common to only one cloud at $T_0$ are identified as a split. Clouds which have been destroyed exist at $T_0$ but have no counterpart at $T_1$, whilst created clouds exist at $T_1$ but have no counterpart at $T_0$.

As mentioned above, for the present we ignore $f_i$ and $g_j$ with $i,j>2$. This is partly motivated by finding that in practice most mergers or splits only involve two clouds (see Section~3.1), and the $f_i$ and $g_j$ with $i,j>2$ tend to be negligible anyway. We find with this assumption the number of merges and splits is reduced by 5-10~\% for $\Delta T=1$ Myr (compared to the total number without the two-body assumption), makes negligible difference for smaller timescales, and leads to about a 20 \% reduction for $\Delta T=5$ Myr. In addition we have not made any constraints about the quantities $f_1, f_2, g_1, g_2$ in the above categories. For example a cloud which has a small but non-zero $f_1$ or $g_1$ is probably not well represented by the `No change' category as it must have undergone a considerable increase or decrease in mass. However this is typically not the case, and the interactions and evolution can largely be considered exclusively from the intercloud medium, except for over long time periods. We could require that `No change' requires $f_1, g_1$ over a certain value, though this would be somewhat arbitrary. Instead we note that, similar to neglecting interactions of $>2$ clouds, there is likely some uncertainty on the fractions of different categories we obtain, again likely a few per cent for $\Delta T=1$ Myr, negligible for lower times, and higher for longer times.

\section{How do clouds evolve?}
\begin{figure}
\centerline{\includegraphics[scale=0.5,  bb=100 180 500 550]{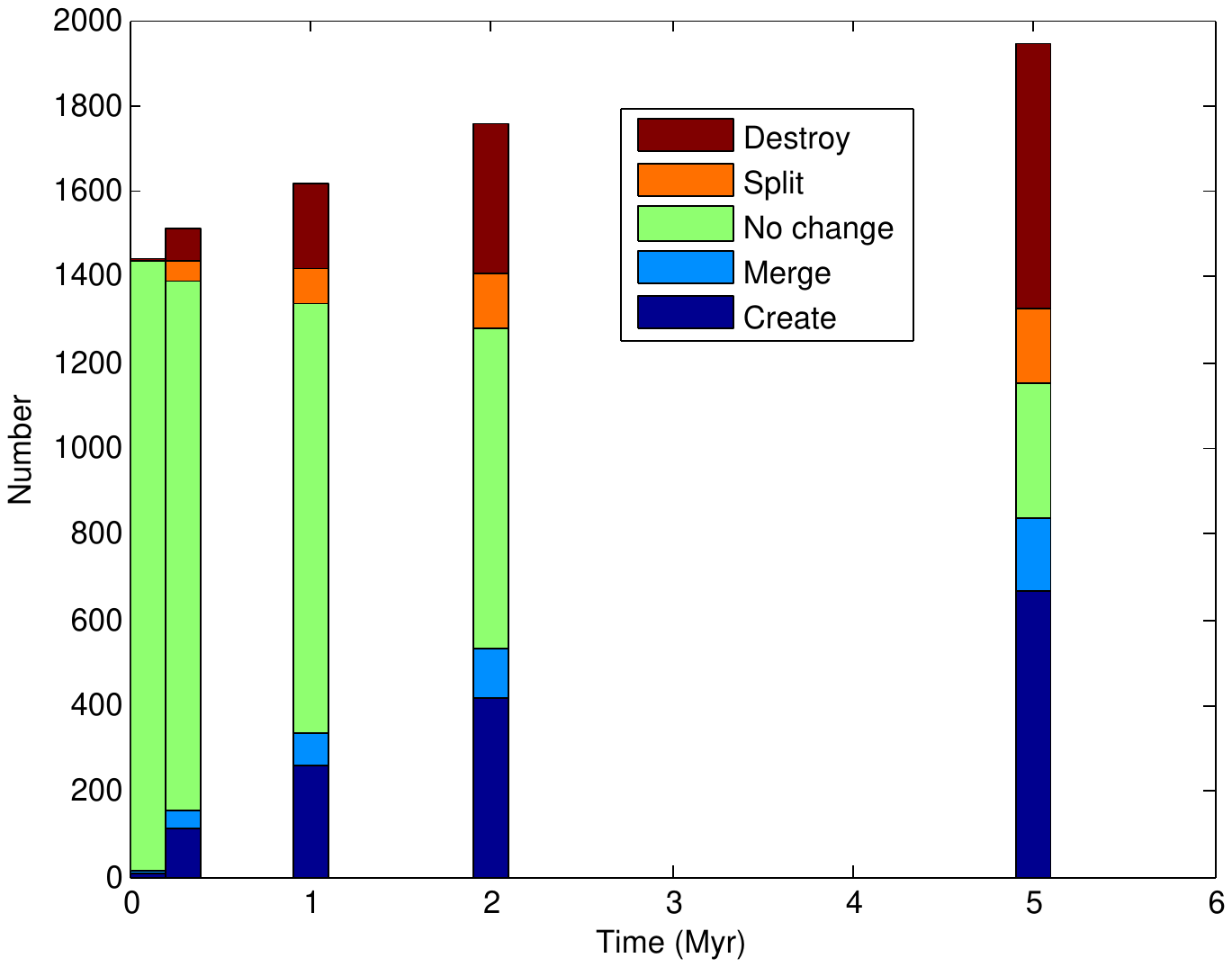}} 
\centerline{\includegraphics[scale=0.5,  bb=100 220 500 520]{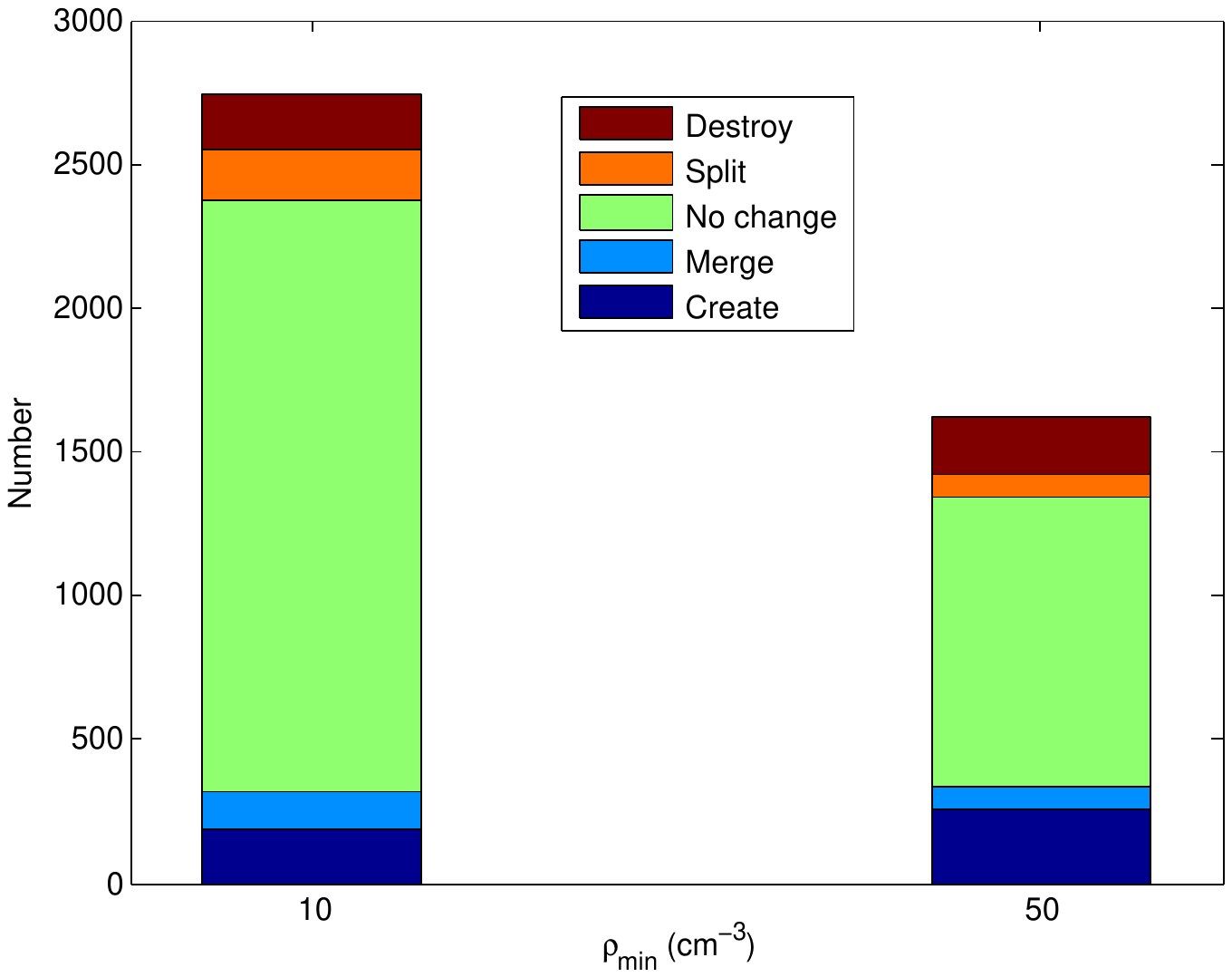}}
\caption{The nature of the evolution of clouds is shown over time periods of 0.1, 0.4, 1,  2 and 5 Myr (top panel), according to the categories defined in Section~2.2. Bars indicate the number of occurrences of each category. Over short timescales, clouds typically undergo no change, as expected. Merges and splits occur in roughly equal numbers and are less frequent than cloud creation or destruction simply from the non-cloud ISM. The lower panel shows the number of clouds in each category when taking $\rho_{min}=10$ and 50 cm$^{-3}$, and $\Delta T=1$ Myr.}
\label{fig:evolution}
\end{figure}

We first show the number of different cloud outcomes according to the categories defined in the previous section. We show these in Figure~\ref{fig:evolution} for different time intervals (top panel) and, different density criteria with a time period of 1 Myr (lower panel). As expected, over a time period of 0.1 Myr, the clouds almost universally undergo no evident evolution. As the time period increases, mergers and splits, cloud creation and cloud destruction become more frequent.  Note that each merger involves two clouds merging, and likewise each split involves one cloud splitting in two.
Generally the number of clouds which are created from the intercloud medium (Create category) and destroyed back to intercloud medium are higher than the number of new clouds spawned from splits, or which disappear through mergers. Figure~\ref{fig:evolution} also demonstrates that typically the number of merges and splits are similar. This could be time dependent, e.g. when cloud growth is predominant at the early stages of the simulation, mergers are likely to be more frequent than splits. At the current stage there is more of a tendency for massive clouds to be broken up, and there is a net increase in the number of clouds, although the system is overall in equilibrium over longer timescales (see \citealt{Dobbs2013} Section~5.1.2 for a discussion of massive clouds forming and dispersing in the arms). The sum of all categories is not consistent because we are including both clouds which are created and destroyed, thus we obtain a sum which is greater than either the number of clouds at $T_0$ or $T_1$ especially over longer time periods. 

As mentioned earlier, for now we only include mergers or splits involving two clouds, leading to an underestimate of the number of mergers. Conversely including any instance of $f_2$ or $g_2>0$, however small, is likely an overestimate. Our assumption of at most two body interactions is less accurate over longer timescales. Over 5 Myr, interactions involving more clouds become more frequent, suggesting that a cloud may have undergone multiple interactions with other clouds (or splits). We tend to focus much of our analysis on timescales of 1 Myr, which appears a reasonable timescale to identify interactions between pairs of clouds, though there is not necessarily a single suitable timescale for all clouds. The duration over which two clouds collide will vary as $l/\sigma_v$ where $l$ is some typical length scale of the cloud and $\sigma_v$ is the cloud-cloud velocity dispersion. We also find interactions of smaller clouds are more frequent. Therefore interactions between smaller clouds (in the sense of both the timescales between interactions and the duration of interactions) may occur over shorter timescales. The timescale of 1 Myr is sufficient to capture most interactions of both large and small clouds.

We also show in Figure~\ref{fig:evolution}  (lower panel) results when using different density criteria for selecting clouds. For the lower density criteria, there are more clouds which undergo no change, splits and mergers. By contrast, the number of clouds created and destroyed is comparatively smaller. This is not so surprising as with the lower density criterion, much more of the gas lies in clouds, so there is less exchange between cloud and intercloud material, and more clouds are formed as the result of splits, or undergo mergers, than form solely from intercloud material.

\subsection{Cloud mergers and splits in more detail}
In this section we consider the values of $f_i$  and $g_j$ for clouds, and thus the validity of our assumption so far that cloud interactions are typically two body processes. 

\begin{figure}
\centerline{\includegraphics[scale=0.5,  bb=100 250 500 540]{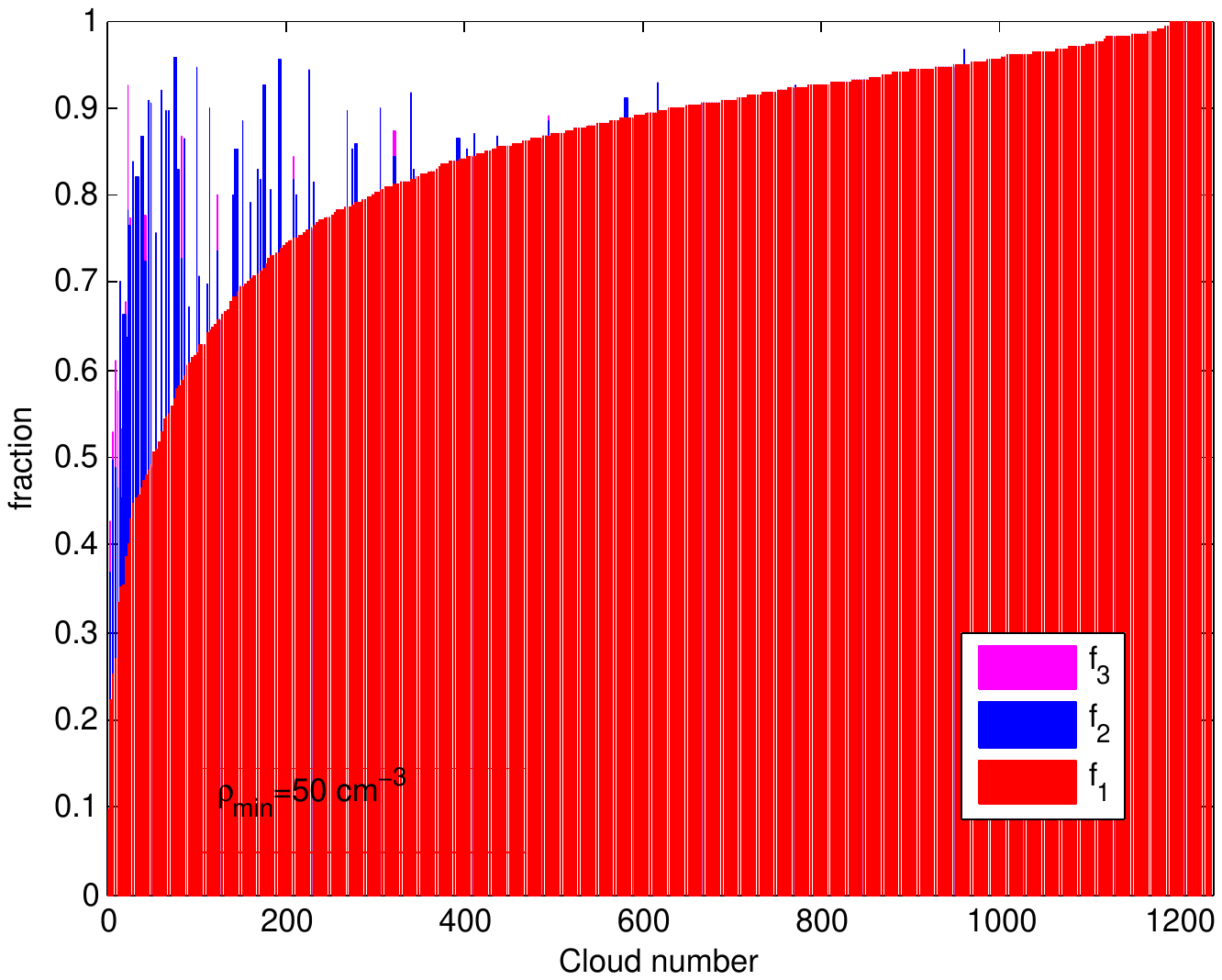}}
\centerline{\includegraphics[scale=0.5,  bb=100 240 500 580]{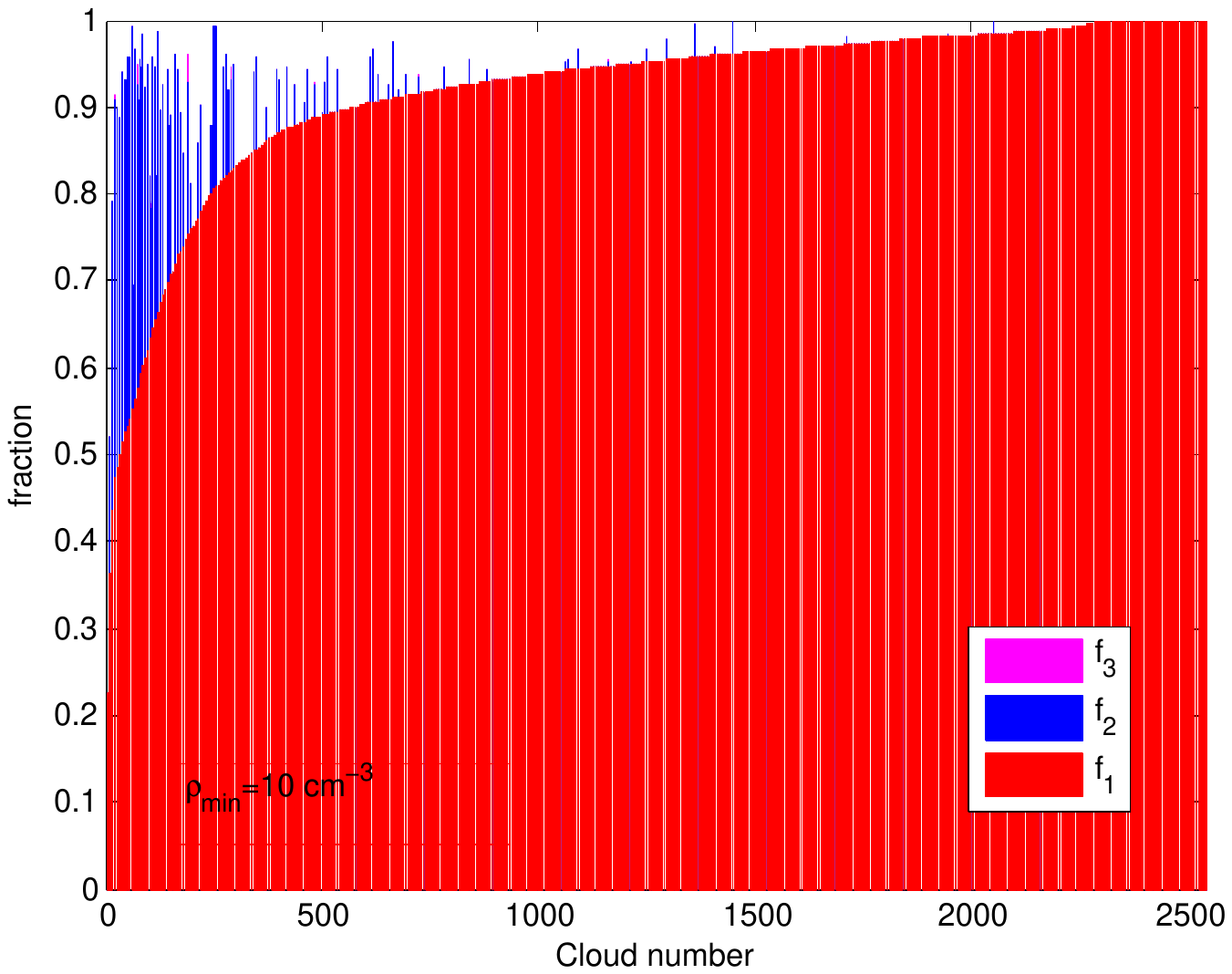}}
\caption{This figure indicates how much gas arises in each cloud at 251 Myr from clouds and intercloud medium over a 1 Myr time period. Clouds at a time of 251 Myr are shown along the $x$ axis. The $y$ axis indicates the fractions of gas in the resultant clouds at T=251 Myr from clouds present at 250 Myr ($f_1,f_2,f_3$ in order of decreasing size) and intercloud medium (the blank space above each column). The panels show the cases where $\rho_{min}=50$ (top) and 10 (lower) cm$^{-3}$.}
\label{fig:f1f2f3}
\end{figure}

In Figure~\ref{fig:f1f2f3} we show $f_1, f_2$ and $f_3$, the fractions of clouds at 250 Myr which now lie in clouds at 251 Myr. The top panel shows results with our fiducial density criteria, the lower panel with the low surface density criteria. Because there are so many clouds, individual bars are not easily distinguishable, but the cases where bars for $f_1$ and $f_2$ are present indicate that a merger has taken place. The blank space above each bar indicates $f_0$. The figure is clearly dominated by the $f_1$'s. This is as expected, as we would expect most clouds to undergo little change over a 1 Myr time interval, and exhibit high $f_1$. As described in the previous sections, the cases with $f_1>0$ but $f_2, f_3=0$ etc. will mostly fall in the `No change' category, and a few in the `Split' category.

There are clear examples in Figure~\ref{fig:f1f2f3} of clouds which lie in the `Merge' category, exhibiting two or three bars indicating non-zero $f_2$, and in a few cases non--zero $f_3$. The number of instances of a non-zero $f_3$ are clearly small. There are only about 10 visible cases in the figure, as indicated by the magenta bars. In most cases, the $f_3$ are also quite small. This justifies assuming that interactions are at most a two body process, as we did for the previous section, at least over a 1Myr timescale. Again, over longer timescales this assumption becomes invalid. We also notice that the mergers are typically on the left hand side of the plot, with lower values of $f_1$. This is again as expected, because we would tend to expect clouds to evolve without significant change in $f_1$, unless they interact with another cloud (or unless those clouds are low density or low mass and near the cloud detection criteria). In the simplest case, of a merger of two equal mass clouds, we would expect $f_1$ and $f_2$ to be $\sim 0.5$. There are mergers similar to this indicated in Figure~\ref{fig:f1f2f3} (with perhaps $f_1,f_2\sim0.4$). There are also cases where $f_1>>f_2$.

In the lower panel of Figure~\ref{fig:f1f2f3} we show $f_1, f_2$ and $f_3$ for our lower density criteria, again over a 1 Myr time period from 250 to 251 Myr. The main difference is that there is less blank space, i.e. $f_0$ is smaller. This again reflects that more of the gas is in clouds, so there is less gas from the intercloud medium which is involved with cloud evolution. This is similar to the finding that with the lower density criteria, mergers and splits are relatively more frequent, whilst created and destroyed clouds are less frequent. For the lower surface density criteria, the $f_3$ are barely visible, indicating again that interactions typically involve only two clouds. Although not evident from the figure, the main exceptions are one or two very massive ($10^7$ M$_{\odot}$) clouds which undergo more mergers with a larger number of small clouds.

\begin{figure}
\centerline{\includegraphics[scale=0.5,  bb=100 250 500 540]{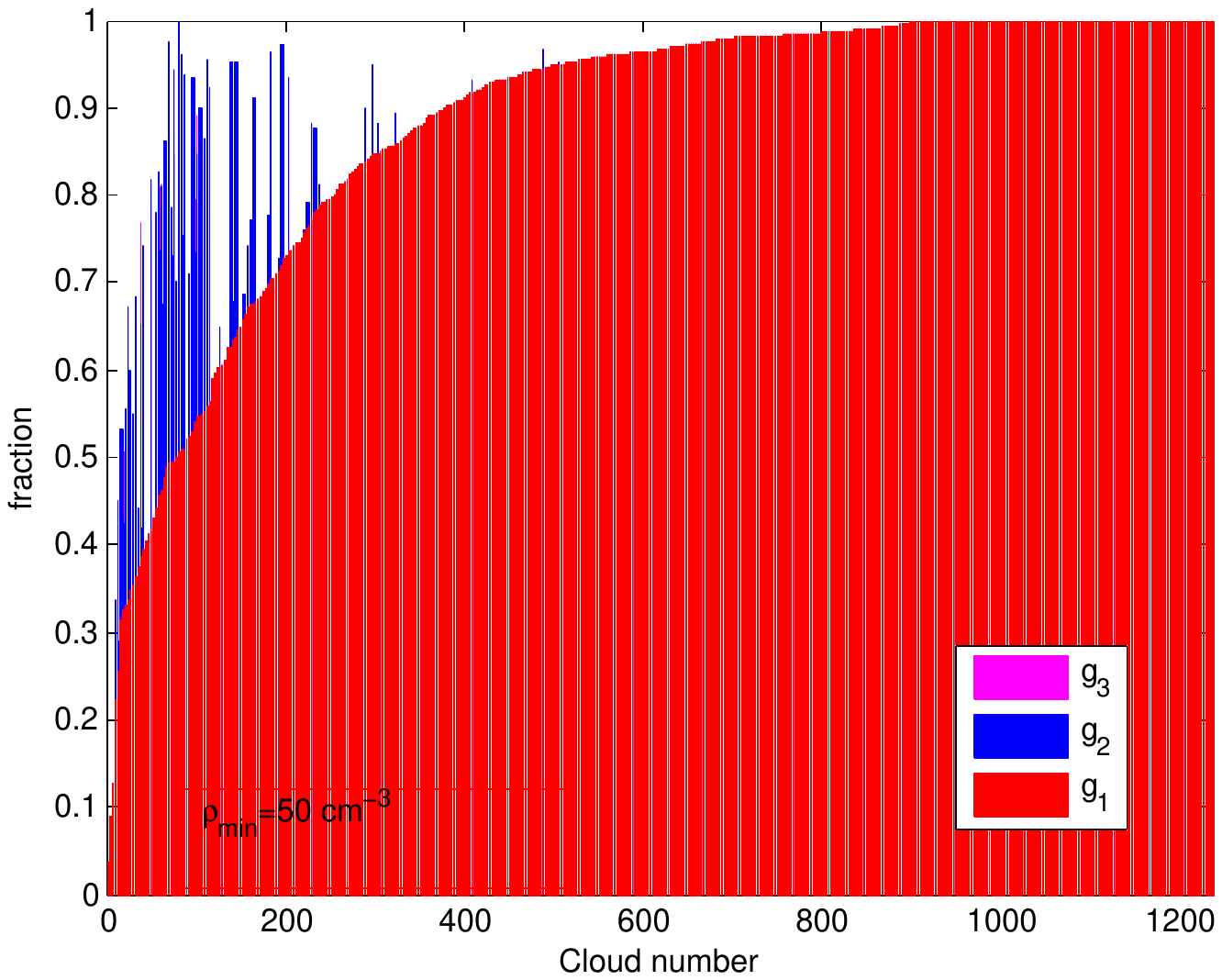}}
\centerline{\includegraphics[scale=0.5,  bb=100 240 500 580]{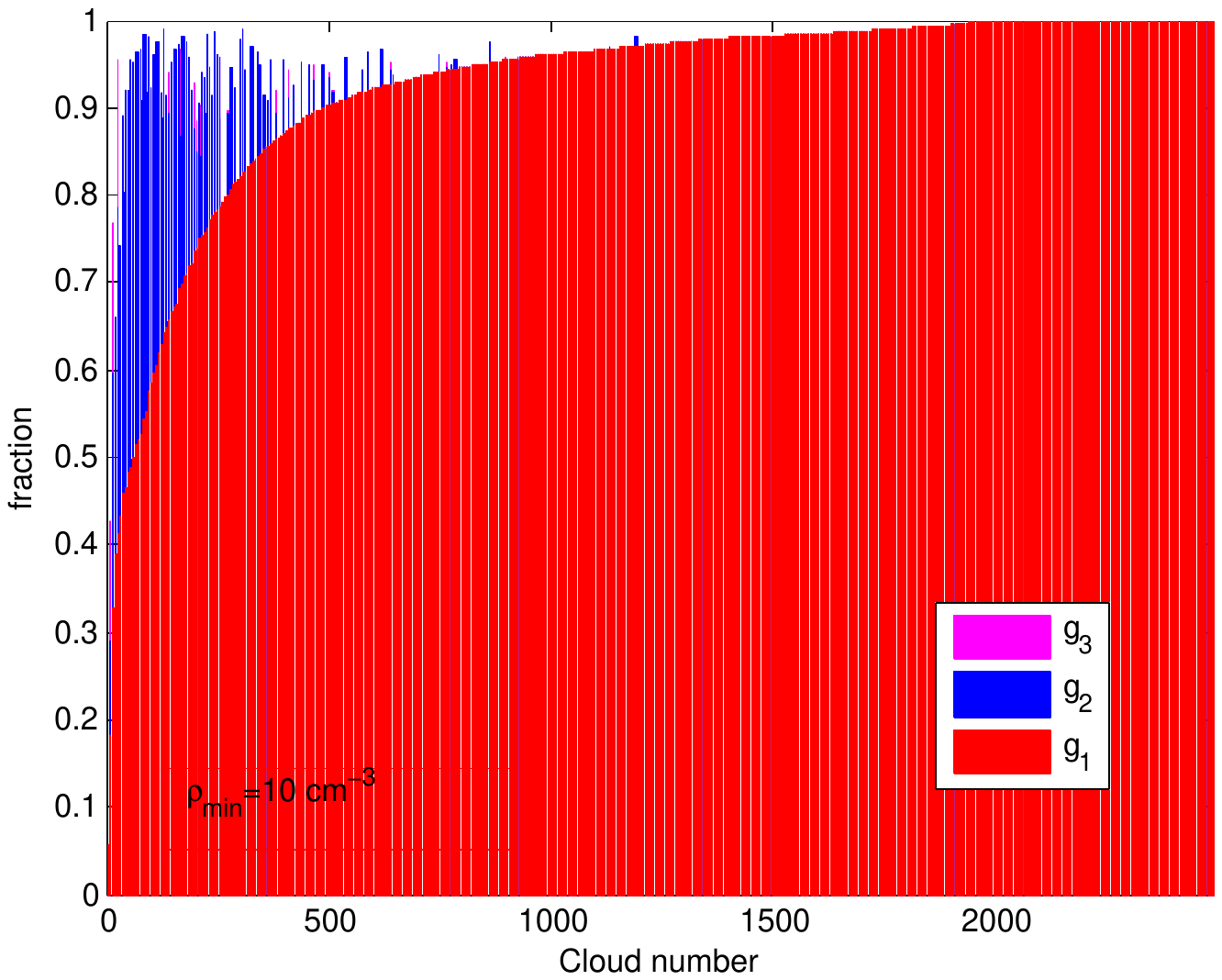}}
\caption{This figure indicates where gas in clouds ends up over a 1 Myr time period. Clouds at a time of 250 Myr are shown along the $x$ axis. The $y$ axis indicates the fractions of gas in these clouds at 250 Myr which ends up in clouds at 251 Myr ($g_1,g_2,g_3$ in order of decreasing size) and intercloud medium (the blank space above each column). The panels show the cases where $\rho_{min}=50$ (top) and 10 (lower) cm$^{-3}$.}
\label{fig:g1g2g3}
\end{figure}
Lastly, in Figure~\ref{fig:g1g2g3} we show the distribution of $g_1, g_2$ and $g_3$, this time indicative of where gas ends up after a 1 Myr time period. 
We again show plots for our standard density criteria (top) and lower density criteria (lower panel). The plots are dominated by $g_1$ indicating that most clouds evolve unperturbed. Clouds which split tend to have lower $g_1$, similar to the $f_1$ for mergers. Again instances with non-zero $g_3$ are rare indicating that clouds typically split up into only two components. 

\subsection{Frequency of cloud--cloud mergers}
As mentioned in the Introduction, the frequency of cloud-cloud collisions is an important diagnostic to determine their relevance to GMC formation in galaxies. Using our analysis, we can estimate the frequency of cloud mergers. We determine the frequency of cloud mergers as 
\begin{equation}
f=\frac{No. \enspace clouds \enspace  involved \enspace  in \enspace  mergers}{Total  \enspace number \enspace  of \enspace  clouds} = \frac{2N_M}{N(T_0)} Myr^{-1}
\end{equation}
over a time period of 1 Myr.
This definition of frequency represents the frequency of mergers experienced by one cloud as it travels around the galaxy (note that in reality the lifetime of the cloud may be less than the time between mergers, see Section~3.6).
We note that it is possible that the clouds can collide and form multiple resultant clouds, e.g. Clouds A and B can merge to produce Clouds C and D. Here we take care not to count the merger of Clouds A and B twice, although the number of instances of this occurring was fairly small. For our fiducial density criterion, if only including two body interactions, we find that 156 clouds are involved in interactions over a 1 Myr period (out of a total of 1442), which gives 0.11 Myr$^{-1}$, roughly 1 every 10 Myr, or 1/14th of an orbit at $R=5$ kpc. Including mergers involving more than two clouds, the frequency is slightly higher, 1 every 8 Myr or 1 every 1/17th of an orbit. We also evaluated the merger frequency for our low surface density criteria, but did not obtain noticeably different values.

\subsection{How much gas arises from cloud vs intercloud medium?}
\begin{figure}
\centerline{\includegraphics[scale=0.45,  bb=100 380 500 750]{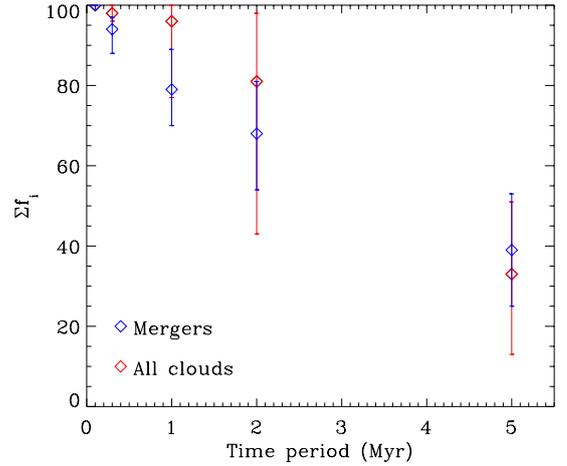}}
\caption{This figure indicates how much gas of clouds existing at T$_1$ was present in clouds at $T_0$, for a given $\Delta T$, described by $\Sigma_i f_i$ (effectively how much gas stays in clouds over a time period $\Delta T$). Results are shown for all clouds (red points) and just those which are the product of mergers (blue points). For timescales of 1 or 2 Myr, gas stays in clouds. However for timescales of 5 Myr, there is much more mixing of cloud and intercloud gas, and above 5 Myr the majority of gas in a cloud originates in intercloud material. The bars indicate the lower and upper quartiles.}
\label{fig:gasinclouds}
\end{figure}

Figures~\ref{fig:f1f2f3}  and \ref{fig:g1g2g3} indicate that generally clouds evolve with little interaction with intercloud medium, at least over 1 Myr. However, we would expect this to change over longer timescales. We show in Figure~\ref{fig:gasinclouds} the median value of the sum of the $f_i$, or equivalently $1-f_0$, over different timescales. We also show separately results for all clouds, and those just for mergers. As expected, minimal gas in clouds originates from the intercloud medium for small $\Delta T$.  Over all clouds, for $\Delta T=1$ Myr, the median value of $f_0$ is 7 \%, or equivalently 93 \% of gas arises from cloud and 7 \% from intercloud material. However by 5 Myr, half of cloud material originates from intercloud material. Thus gas is being exchanged in clouds over a timescale of several Myrs, in agreement with Dobbs \& Pringle 2013 where we found GMC lifetimes of order several Myrs. The sum of $f_i$ tend to be smaller for mergers compared to cloud evolution generally, reflecting that $f_1$ is typically lower for mergers (Figure~\ref{fig:f1f2f3}) and mergers tend to involve more disruption. For the longer timescale, the opposite is true, i.e. mergers reflect a lower fraction of intercloud material, perhaps indicating that mergers are sustaining clouds over longer time periods.  As expected the timescales for gas in clouds and intercloud gas to exchange is longer with the lower density criterion (not plotted), in this case the amount of gas retained in clouds after 5 Myr ($1-f_0$) is still over 70 \%. Again this reflects similar findings by \citet{Dobbs2013} that Giant Molecular Associations (GMAs) are likely to have longer lifetimes.

\subsection{Typical values of $f_1$ and $f_2$}
\begin{figure}
\centerline{\includegraphics[scale=0.5,  bb=100 220 500 540]{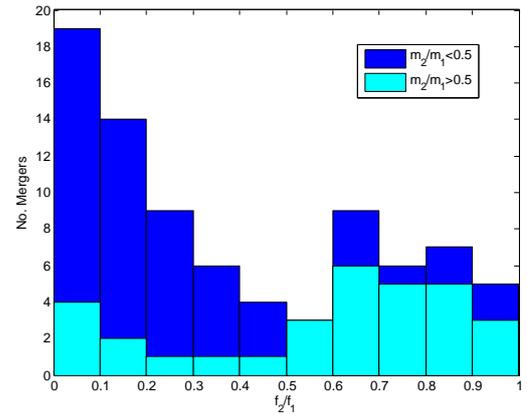}}
\caption{The number of mergers is shown for different ratios of $f_2/f_1$, so for example $f_2/f_1\sim1$ indicates roughly equal amounts of mass originate from merging clouds. The clouds are further classified by the ratio of the cloud masses of the merging clouds, $m_2/m_1$.}
\label{fig:gasinmergers}
\end{figure}

Figure~\ref{fig:f1f2f3} showed mergers of 2 or more clouds, but it is difficult to tell whether mergers are typically between clouds of equal mass, or smaller clouds joining larger clouds. In Figure~\ref{fig:gasinmergers} we show the numbers of mergers for different ratios of $f_2/f_1$, again with $\Delta T=$ 1 Myr. Overall Figure~\ref{fig:gasinmergers} indicates that the majority of mergers have small $f_2$ compared to $f_1$.
If $f_2/f_1<<1$, this implies either that a small cloud has merged with a much larger cloud, or that two more equal mass clouds have collided, but only a small fraction of one has merged with the other. The latter could arise from a  grazing, or offset collision of two clouds. To differentiate between these two scenarios, we have divided the mergers in Figure~\ref{fig:gasinmergers} into two populations depending on the original mass of the clouds colliding. We divide mergers depending on whether the ratio of the original masses, denoted $m_2/m_1$, is greater or less than 0.5. If $m_2/m_1>0.5$ then the clouds merging had similar masses. Figure~\ref{fig:gasinmergers} indicates that for those mergers with $f_2/f_1<<1$, typically $m_2/m_1<0.5$ indicating that a small cloud is merging with a large cloud. There are relatively few cases of two similar mass clouds interacting, but only a small transfer of mass from one to the other. Similarly most the mergers with $f_2/f_1\sim1$ involve two similar mass clouds, rather than an unequal mass transfer during the merger.

Again we may expect the values of $f_1$ and $f_2$, and the nature of mergers, to be time dependent. For example a grazing collision of two clouds, with only a small transfer of mass from one cloud to another, would be indistinguishable from a full merger of a smaller cloud with a more massive cloud, when viewed over a short time period. To test this, we looked at results for $\Delta T=5$ Myr (not shown in Figure~\ref{fig:gasinmergers}). In this case, a higher fraction (around $\sim 40$ \%) of clouds with small $f_2/f_1$ have more equal masses ($m_2/m_1\sim1$), indicative of grazing collisions, though the majority still represent the full merger of a smaller cloud. 

\begin{figure}
\centerline{\includegraphics[scale=0.5,  bb=100 250 500 540]{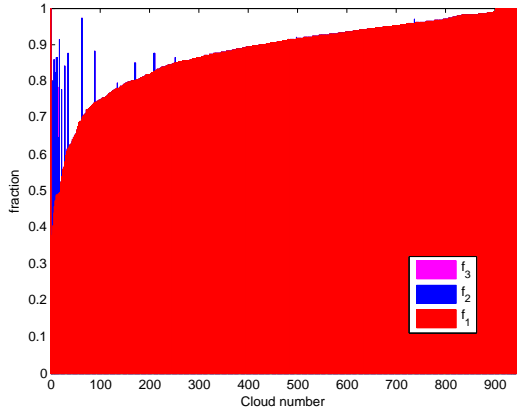}}
\caption{The values of $f_1,f_2$ and $f_3$ are shown over a 1 Myr time period, for a simulation with no imposed spiral arms. These represent the fraction of gas in clouds at a time of 251 Myr, which was present in clouds at 250 Myr. Clouds at a time of 251 Myr are shown along the $x$ axis. The $y$ axis indicates the fractions of gas $f_1, f_2$ and $f_3$, in order of decreasing size, and intercloud medium (the blank space above each column). The standard density criteria, with $\rho_{min}=50$ cm$^{-3}$ is used. The main difference compared to Figure~\ref{fig:f1f2f3} is that there are fewer blue bars present (for $f_2$) indicating fewer mergers in the absence of spiral arms.}
\label{fig:f1f2f3noarm}
\end{figure}

\subsection{Frequency of mergers without spiral arms}
We also investigate cloud evolution in galaxies without a strong spiral arm perturbation, and in particular the frequency of cloud mergers. In this section we analyse results from a simulation with no imposed spiral potential. We find that over a 1 Myr period, 42 out of 1184 clouds are involved in interactions with each other. This gives a frequency of 0.035 Myr$^{-1}$, once every 28 Myr, or once every 1/5 an orbit at a radius of 5 kpc . As expected, the frequency of mergers is notably less compared to the simulation with spiral arms. However the interactions are still relatively frequent compared with theoretical estimates (see Introduction and Section~5). This is because even without an imposed spiral potential, the gas gathers into dense, sheared features like short sections of spiral arm, simply due to self--gravity and thermal instabilities (see Figure~\ref{fig:columndensity}). Hence, clouds are still concentrated into spiral-like features, rather than being spread uniformly over the disc.  

We also determined $f_0, f_1, f_2$ and $f_3$ as for Section~3.1, for the case without spiral arms (see Figure~\ref{fig:f1f2f3noarm}). Again nearly all interactions only involved two clouds. We found the median value of $f_0$ was 9 \%, indicating 91 \% of gas stays in clouds compared to 9 \% from the intercloud medium. These fractions are quite similar, just a slightly higher $f_0$, compared to the case with spiral arms. 

\subsection{Comparison of cloud merger rates with cloud lifetimes}
In \citet{Dobbs2013}, we analysed cloud lifetimes and concluded that most clouds have quite short lifetimes $\lesssim 10$ Myr. Thus for clouds in the spiral arms, we would expect most to only experience one or two mergers during their lifetime. However in \citet{Dobbs2013} we also noted that some clouds, including a number of more massive $10^6$ M$_{\odot}$ clouds, exhibited lifetimes of 20 Myr or more, and thus could undergo multiple mergers over their lifetime. Fundamentally we expect the lifetimes to be a result of clouds merging, since this builds up mass into more massive and longer lived clouds. In the spiral arms, clouds are stochastically able to undergo multiple collisions, before stochastically being destroyed. Conversely in the simulations with no spiral arms, the cloud merger rate is small, at least for $>10^4$ M$_{\odot}$ clouds. Thus the probability of acquiring a high mass, long lived cloud is low.

\section{Mergers of massive clouds}
As mentioned in the Introduction, collisions of more massive clouds may be of particular interest as they may be associated with massive star clusters. In particular, collisions may enable a large mass of gas to be delivered to a massive, dense GMC in a short period of time, thus accounting for small age spreads in massive clusters \citep{Furukawa2009,Fukui2014}. In this section we present and discuss the examples of mergers of massive clouds which are occurring at our chosen time frame of 250--251 Myr, in the simulation with an imposed spiral potential.
\begin{figure}
\centerline{\includegraphics[scale=0.4]{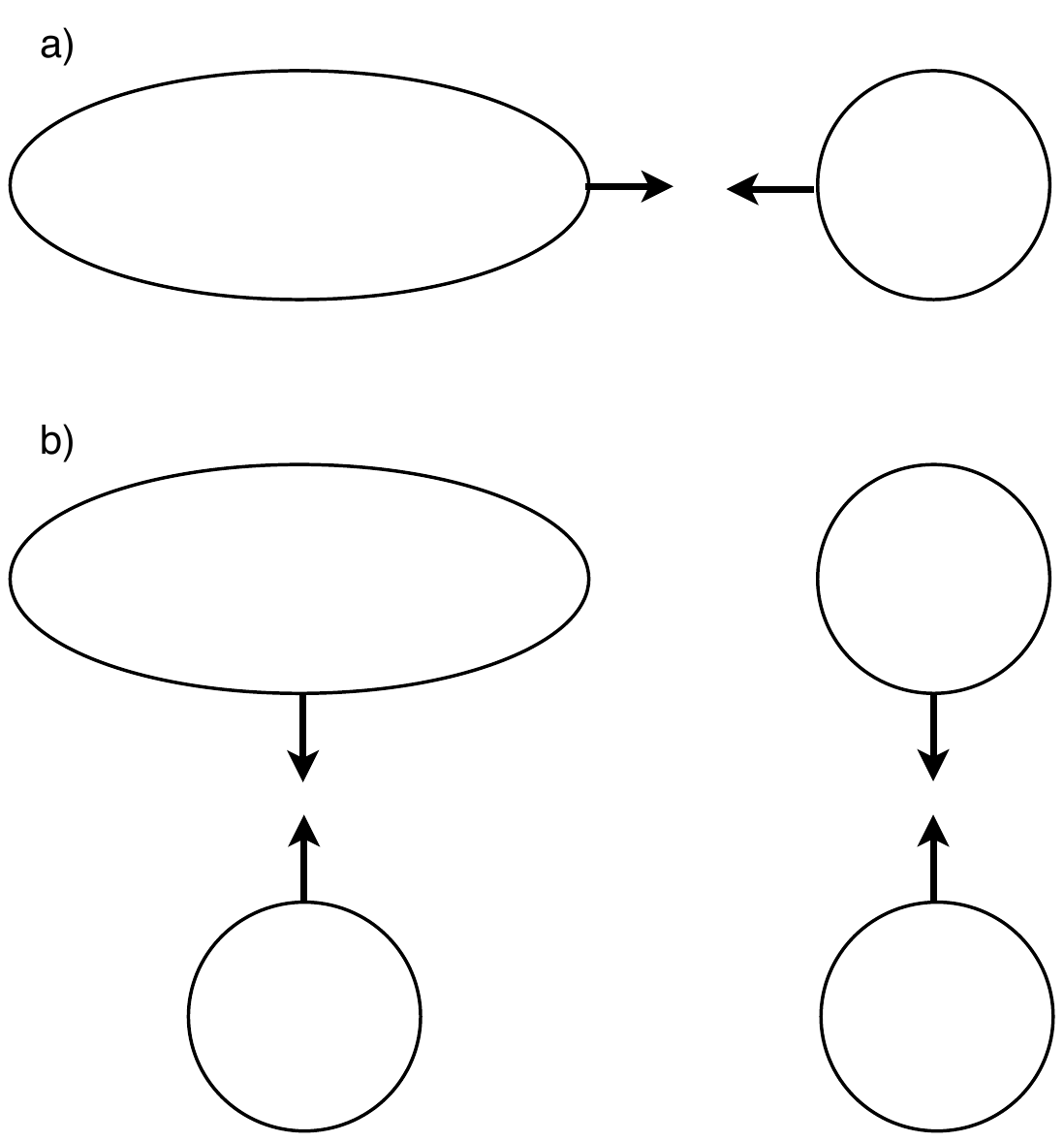}}
\caption{Possible orientations for mergers, or collisions, are shown. A `filamentary' merger (a) is characterised by the intersection of the clouds along the minor axis of the largest cloud. Examples of `full on' mergers (b) are characterised by intersection of the clouds along the major axis of the largest cloud, or where neither of the clouds are particularly elongated (in this paper elongated clouds have aspect ratio $>$2).}
\label{fig:cartoon}
\end{figure}

\begin{figure}
\centerline{\includegraphics[scale=0.4,   bb=100 170 500 480]{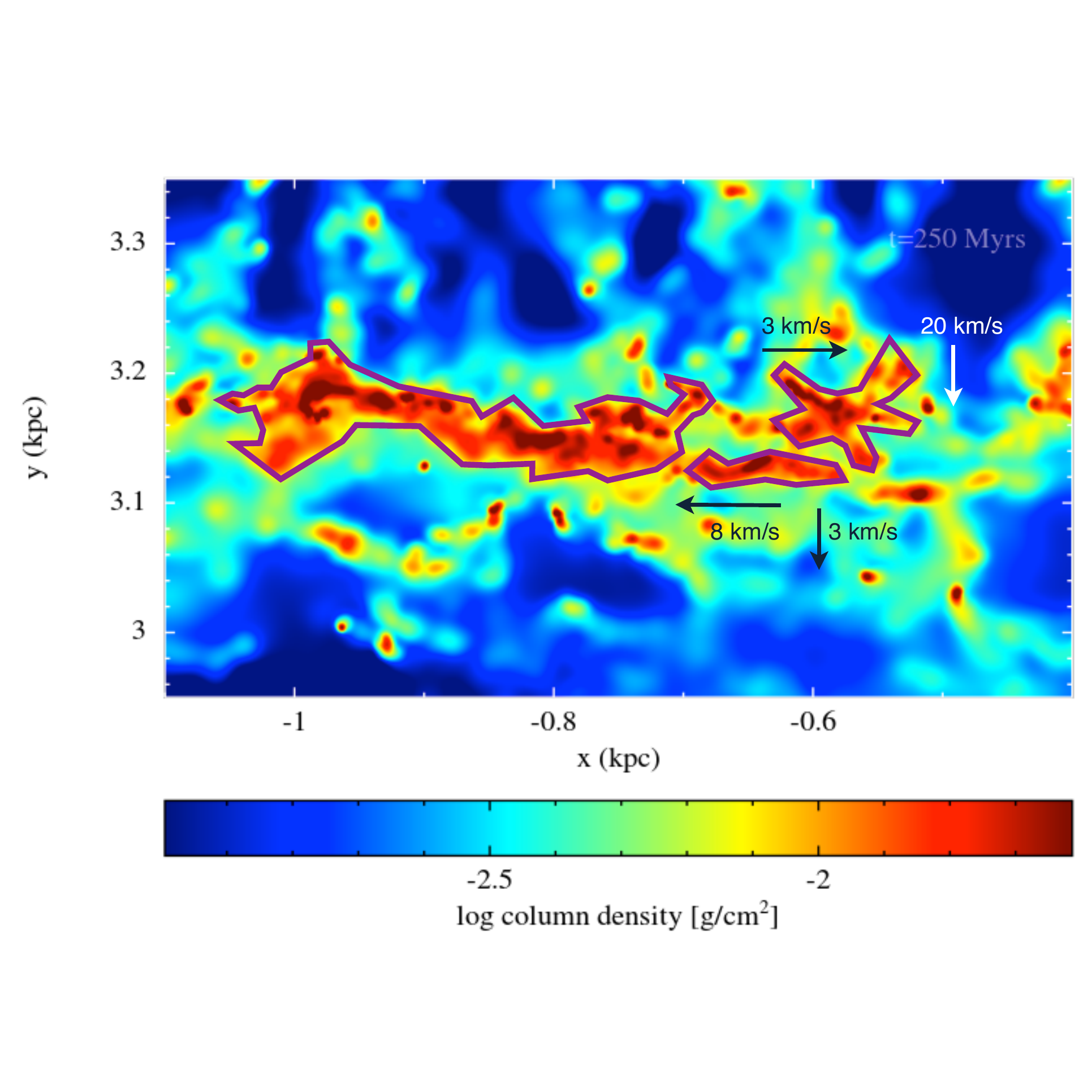}}
\centerline{\includegraphics[scale=0.42,   bb=70 0 500 550]{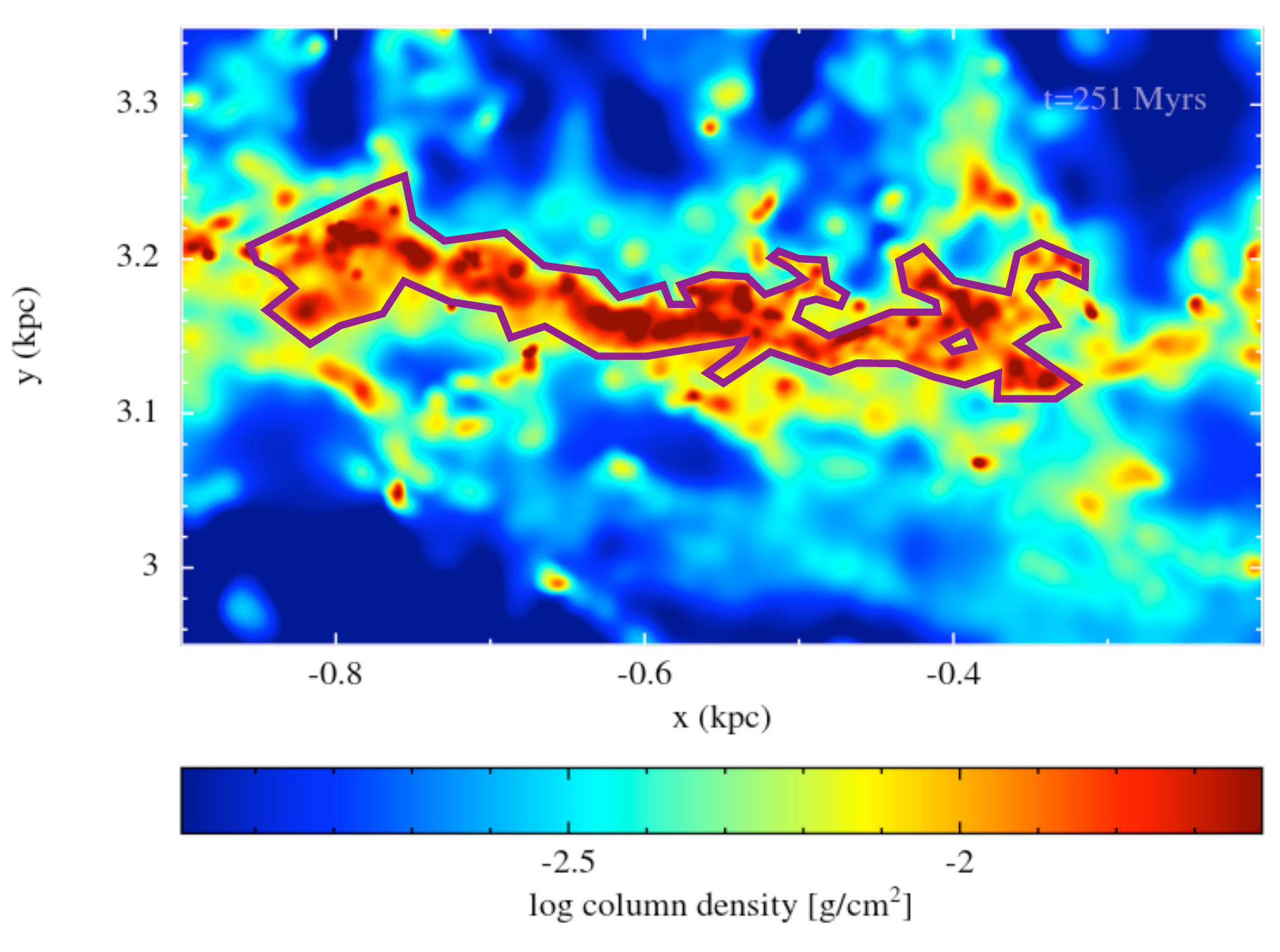}}
\caption{`Merger 1'  (from Table~2) involving three clouds is shown at 250 Myr (top), before the clouds merge, and at 251 Myr (lower), when the three clouds have merged into a more massive cloud. Details of the masses of the clouds are given in Table~2.  Arrows indicate the relative velocities of clouds compared to the left hand cloud. The boundaries of the clouds found by the clump-finding algorithm are indicated by the thick purple lines. The boundaries are drawn by hand -- an automated method was tried first, but the hand drawn boundaries gave more successful indications of the shapes of the clouds compared to an automated method.}
\label{fig:collision1}
\end{figure}

\begin{figure}
\centerline{\includegraphics[scale=0.28]{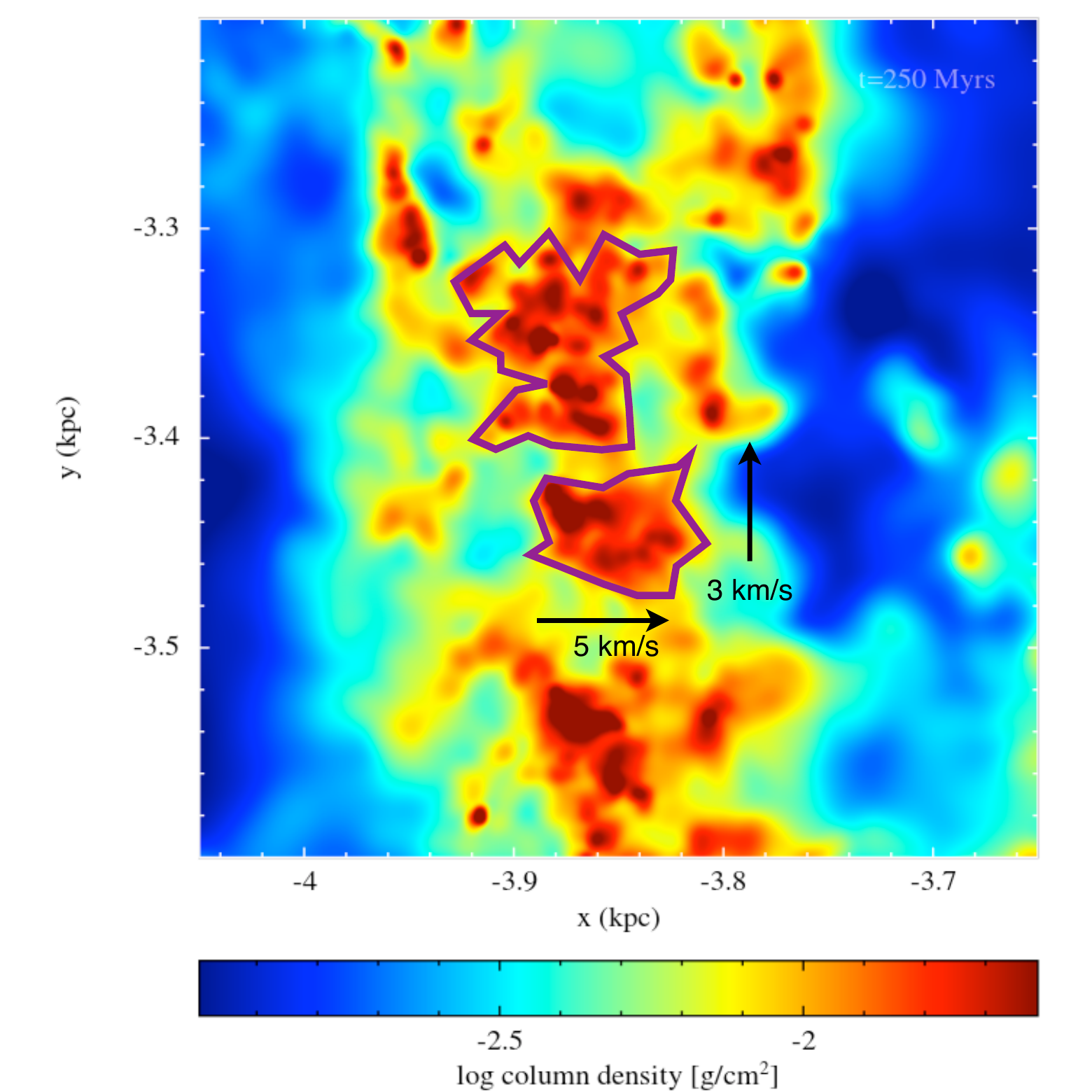}
\includegraphics[scale=0.28]{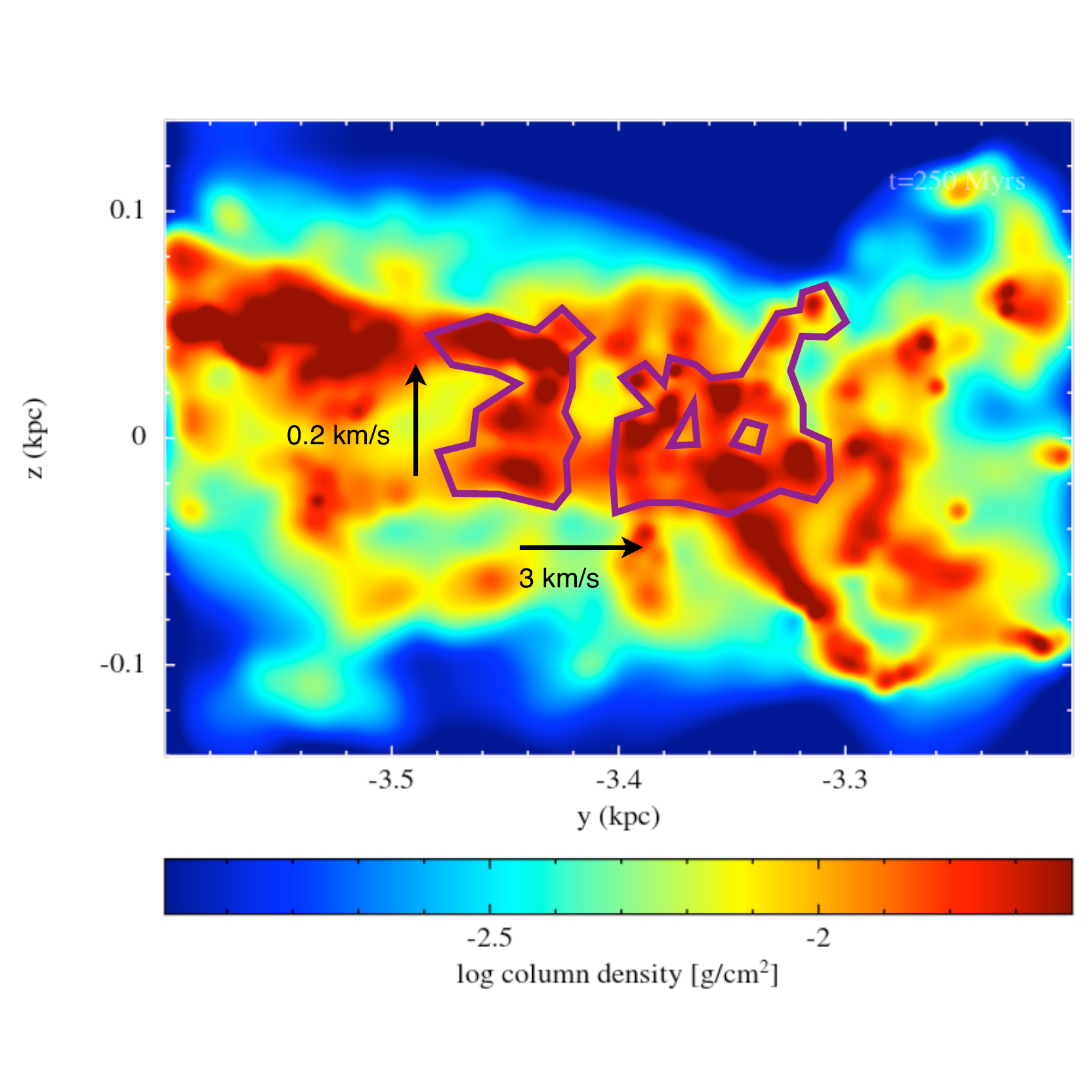}}
\centerline{\includegraphics[scale=0.28]{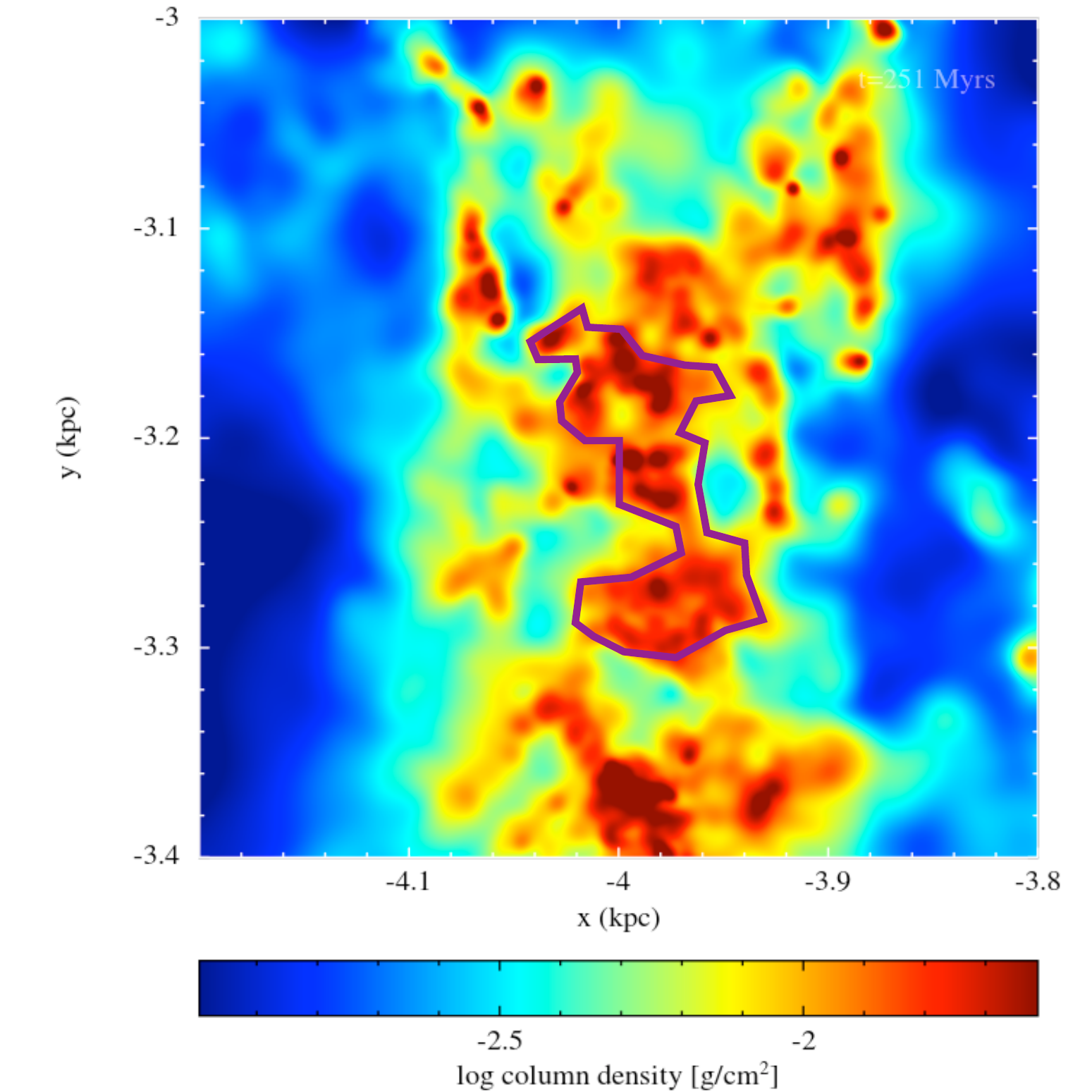}
\includegraphics[scale=0.28]{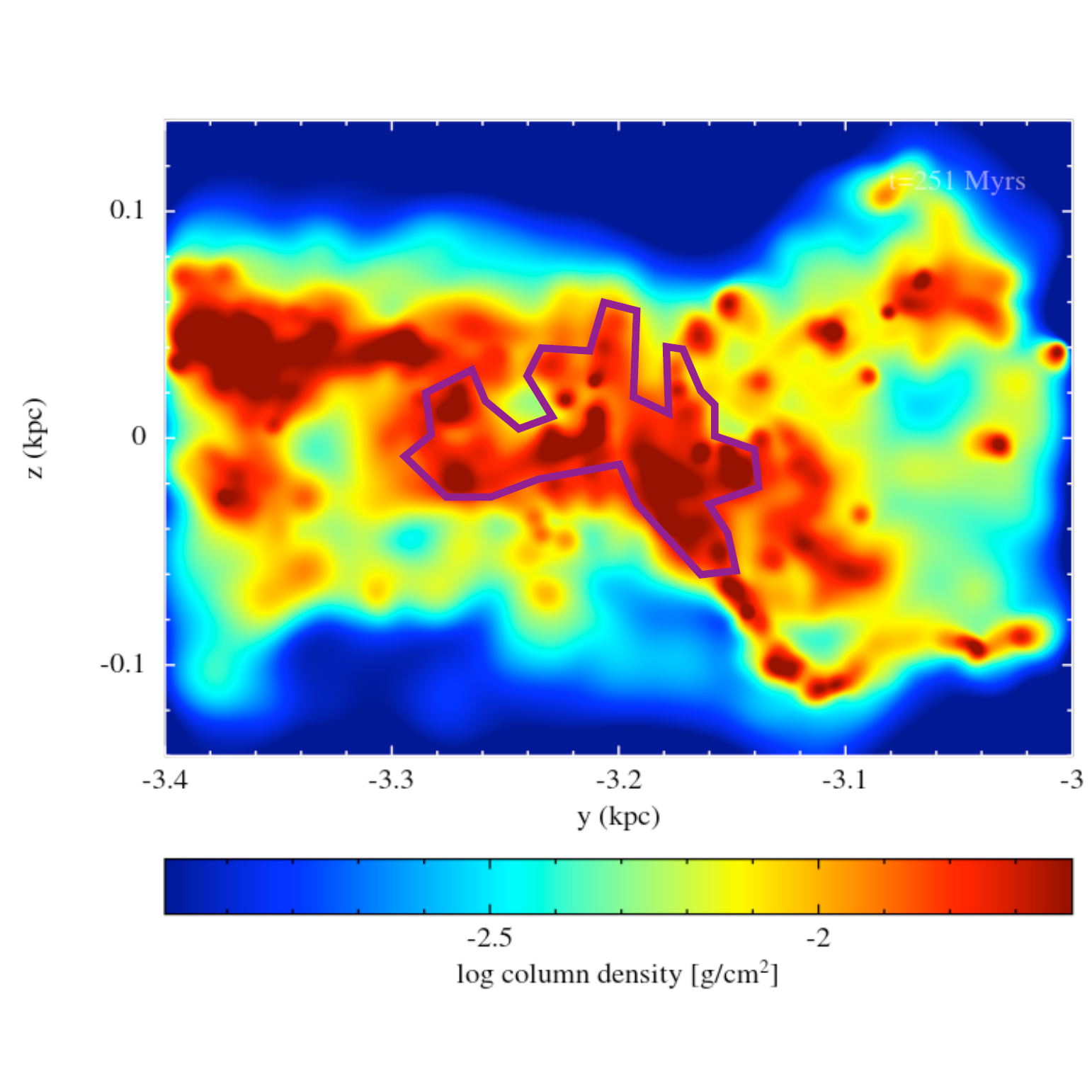}}
\caption{`Merger 2'  (from Table~2) is shown at 250 Myr (top), before the clouds merge, and at 251 Myr (lower), after the interaction. The left hand plots show the $xy$ (face on) view and the right hand plots the $yz$ (edge on) view. Details of the masses of the clouds are given in Table~2.  Arrows indicate the relative velocities of the lower / left hand cloud compared to that on the top / right. This merger is also shown in H$_2$ and CO in Figures~\ref{fig:h2} and \ref{fig:co}.}
\label{fig:collision2}
\end{figure}

We first considered whether taking a higher mass cut off changed our results in the earlier parts of the paper (how the clouds evolve, and the fractions $f_0, f_1, f_2$), but we did not find significant differences. We then calculated the frequency of mergers of more massive clouds. Over a 1 Myr time period, taking $\rho_{min}=50$~cm$^{-3}$ (this is the total, atomic plus molecular density) to select clouds, we find seven mergers of two or more clouds with masses $>10^5$~M$_{\odot}$, a frequency of 0.0049 Myr$^{-1}$, or one merger every 206 Myr (one every 1 1/2 orbital periods). The details of the mergers are shown in Table~2, where we can see that one case involves three clouds merging to form one cloud, the other cases only involve two clouds. The last column indicates whether the interaction is filamentary in nature (see Figure~\ref{fig:cartoon}). Because clouds are sometimes elongated, and furthermore roughly aligned with the spiral arms, mergers of this type are surprisingly frequent (compared to a distribution of randomly orientated clouds). Mergers not labelled as `filamentary' involve more full on collisions between typically less elongated clouds. The majority of our examples are `filamentary' rather than  `full on' in nature.
\begin{table*}
\centering
\begin{tabular}{c|cc|c|c|c|c}
 \hline 
Merger & Mass of 1st cloud & Mass of 2nd cloud & Mass of 3rd cloud & Mass of resultant & Filamentary? & Merger velocity \\
& ($10^5$) M$_{\odot}$ & ($10^5$ M$_{\odot}$) & ($10^5$ M$_{\odot}$) & cloud ($10^5$ M$_{\odot}$)  & & (km s$^{-1}$) \\
 \hline
1 & 1.5 & 5.5 & 1.1 &10.0 & Y & 17 and 8 \\
2 & 1.8 & 1.4 & - & 1.3 & N & 3 \\
3 & 10.3 & 3.3 & - & 15.1 & Y & 8 \\
4 & 16.3 & 1.0 & - & 9.2 & Y & 2 \\
5 & 2.2 & 2.0 & - & 4.0 & N & 4 \\
6 & 1.1 & 1.1 & - & 2.4 & Y & 9 \\
7 & 23 & 1.8 & - & 2.7 & Y & 8 \\
\hline
\end{tabular}
\caption{Table showing the details of mergers between massive clouds. Mergers 1 and 2 are shown in Figures~\ref{fig:collision1} and \ref{fig:collision2}. The merger velocities are calculated as the relative velocities between the two clouds (for Merger 1 two pairs of clouds are colliding). The mergers occur predominantly in the plane of the disc, except for Merger 4.}
\label{tab:runs}
\end{table*}

Figure~\ref{fig:collision1} shows Merger 1, which involves one cloud of $5\times10^5$ M$_{\odot}$ and two clouds of $\sim10^5$ M$_{\odot}$. The most massive cloud is elongated, aligned with the spiral arm. The two smaller clouds merge onto the most massive cloud at one end. From the velocities, it seems the far right cloud is moving downwards onto the centre cloud, and the centre cloud is merging with the more massive, left hand, cloud. However the far right cloud is actually moving away from the far left cloud, so the three clouds do not appear to be totally convergent. The merger of the three clouds results in an even further elongated cloud (lower panel), with perhaps just a small increase in dense gas in the centre. Thus the collision does not appear to have a strong impact on the structure of the clouds. The interaction in Figure~\ref{fig:collision1} highlights the difficulty in proposing the collision of massive clouds to produce large amounts of dense gas quickly. Many of the clouds are elongated and aligned with spiral arms, so tend to interact along their minor axes (corresponding to Figure~\ref{fig:cartoon}, top panel). Thus they collide only over a small cross-section. The  density structure of the resultant cloud also suggests that rather than a violent collision, the smaller clouds gently merge onto the end of the larger cloud. The clouds could potentially reach high densities where they collide but the collision interface not be fully resolved by the simulations, however the geometry of the collision suggests that this would be limited to localised areas, and would need higher resolution simulations to study in detail. The velocities of the clouds are relatively high, compared to the other examples (see Table~2, Figure~\ref{fig:collision2}  and Figure~\ref{fig:dispersions}). This could be conducive to triggering star formation -- simulations of isolated collisions have suggested that higher Mach numbers produce a stronger shock compressed layer, leading to higher rates of star formation \citep{Bekki2004,Kitsionas2007}. However the relative velocities tend to be higher for the filamentary collisions partly because the biggest cloud simply covers a relatively large area spanning a wider range of velocities. 

Figure~\ref{fig:collision2} shows a second example of a merger, both top down and edge on. In the edge on view the clouds appear to merge more fully compared to the top down view. Overall, similar to the collision in Figure~\ref{fig:collision1}, there is no particular indication that the cloud density structure is changed by the interaction. Rather the interactions resemble two clouds coming together, but their own structures remaining fairly similar. The relative velocity of the clouds is fairly small, indicating only a mildly supersonic collision. Merger 5 (not shown in any figures) is the only other example which is not one cloud joining the end of another. The clouds in this case appear more intricately linked before the collision. Unusually, the clouds are in an inter-arm spur. Whilst this does not necessarily rule out massive cluster formation, it is counter-intuitive since these clouds are likely at the end of their lifetime. Also these clouds tend to have a large stellar age spread \citep{Dobbs2014}.

Our simulations highlight the difficulties of producing massive clusters via collisions when considering that clouds are often elongated and collisions often represent the addition of a smaller cloud at the end of one of these clouds. One possible limitation is that we have used a relatively low surface density of 8 M$_{\odot}$ pc$^{-2}$. Hence we also analysed cloud mergers in a simulation with a surface density of 16 M$_{\odot}$ pc$^{-2}$ (see Section~2), more similar to the inner Galaxy. However we tended to find not much difference from the mergers shown in Table~2, with most still involving somewhat elongated clouds, and little apparent change in cloud structure due to the collision. Another limitation is the maximum density imposed by the inclusion of feedback, and that we cannot resolve the shoctk at the interface of the collision, but much higher resolution simulations would be needed to investigate this. \citet{Inoue2013} also suggest that magnetic fields enhanced by shocks could promote the formation of massive cores, but again this is far beyond the resolution of galactic scale simulations.

\section{Theoretical Comparison}
As discussed in the introduction, cloud-cloud collisions have been the focus of a number of theoretical studies. We compare here the frequency of mergers found in our simulations with results expected from analytic estimates. To calculate collision, or merger frequencies, we need the cloud-cloud velocity dispersion, so we calculate that next. 

\subsection{Cloud--cloud velocity dispersions}
A number of works have considered the internal velocity dispersions of clouds in galactic simulations \citep{Tasker2009,Dobbs2011new}, but here we determine cloud-cloud velocity dispersions. As well as being a factor which governs the frequency, and possibly nature of cloud--cloud collisions, we can also compare our cloud--cloud velocity dispersions with observations.

\begin{figure}
\centerline{\includegraphics[scale=0.43]{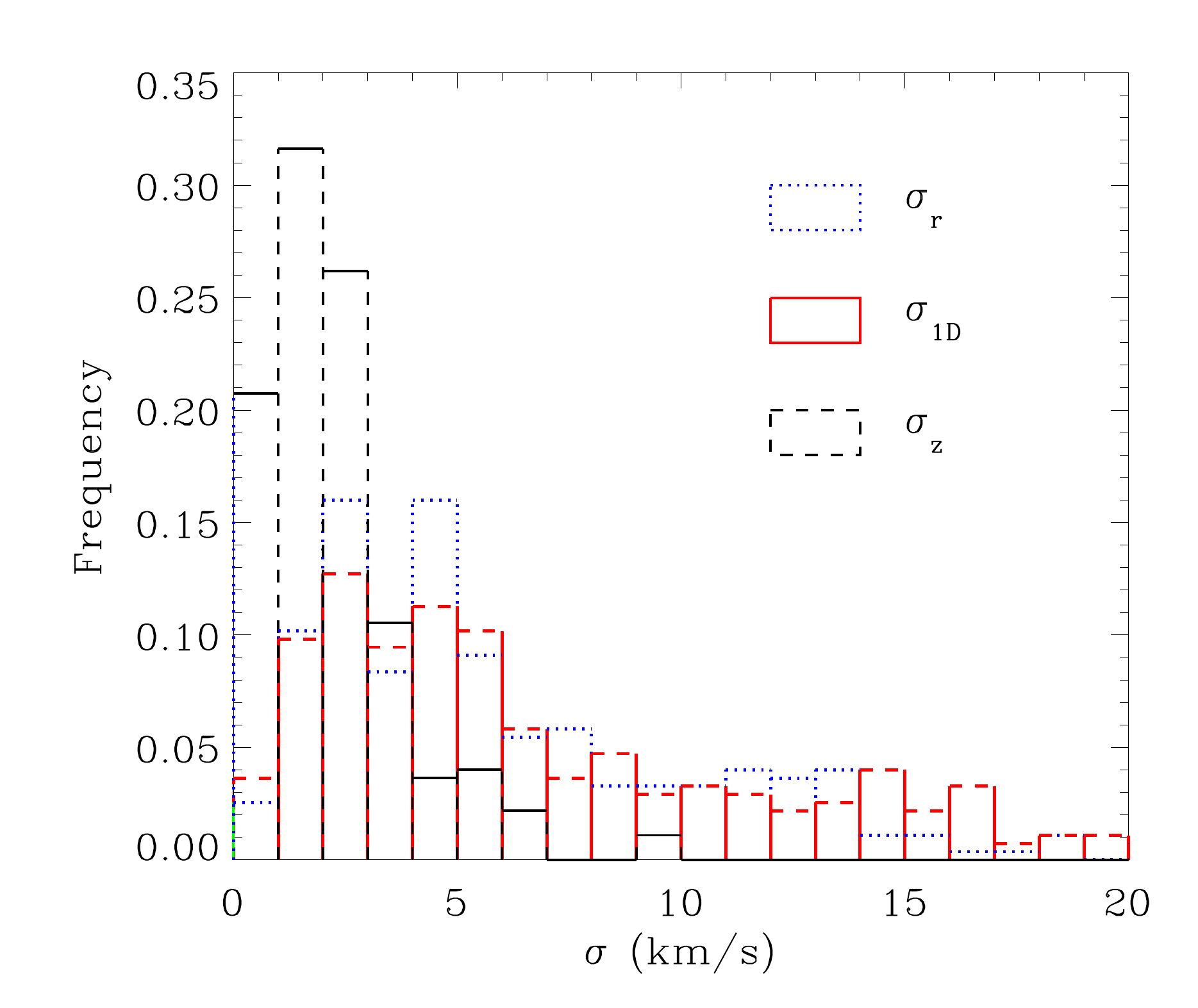}}
\caption{Cloud-cloud velocity dispersions are shown for an average 1 D dispersion, the dispersion in the plane of the disc, and that in the vertical direction. The dispersions tend to be at the lower end of observed values, the median dispersion either averaged in 1D or in the plane of the disc being around 4-5 km s$^{-1}$.}
\label{fig:dispersions}
\end{figure}

Cloud-cloud velocity dispersions are shown in Figure~\ref{fig:dispersions}. The velocity dispersions are calculated using the velocities of clouds within regions of dimensions of 500 pc by 500 pc. Taking smaller or larger regions shifted the dispersions to slightly ($\sim$ 10\%) lower and higher values respectively, although once the size scale reached $\lesssim 100$ pc there were too few clouds to properly compute dispersions. We computed the dispersion in the plane of the disc ($\sigma_r$), the mean 1D dispersion ($\sigma_{1D}$), and the dispersion in the $z$ direction ($\sigma_z$). We find that in the plane of the disc the 1D dispersions are around 3--6 km s$^{-1}$. By comparison, \citet{Wilson2011} find an average cloud-cloud velocity dispersion of 6.1 km s$^{-1}$ from a sample of 9 galaxies, whilst a similar dispersion is found for the Milky Way clouds \citep{Stark2005}. Our 1D dispersions, which can be compared to the 1D dispersions found by \citet{Wilson2011}, are generally smaller than the observations, but do compare favourably with some galaxies with low dispersions such as NGC 628. We also find that cloud-cloud dispersions are lower in the vertical direction, compared to the plane of the disc. The relatively low values of the velocity dispersions also suggest that most cloud--cloud interactions will not be particularly disruptive. We also note that with a velocity dispersion of 6 km s$^{-1}$, a cloud will travel only about 6 pc per Myr, so short--lived clouds will only interact with other clouds formed in a close proximity. 

We also examined the cloud-cloud velocity dispersions in our model without imposed spiral arms (not plotted). The distribution of $\sigma_r$, $\sigma_{1D}$, and $\sigma_z$ are similar but shifted roughly 1 km s$^{-1}$ lower. This could be because the spiral arms generally introduce slightly larger velocity dispersions in the gas, due to spiral shocks \citep{Bonnell2006,Dobbs2007a,Dobbs2011new}. There is also less difference between the velocity dispersion in the vertical direction compared with the other velocity dispersions.

\subsection{Cloud-cloud collision rates}
The time between collisions can be estimated theoretically from the cloud-cloud mean free path. If we assume the clouds are spherical (and ignore gravity) then the time between collisions is expected to be 
\begin{equation}
t_{coll}=\frac{\lambda}{v}  = \frac{1}{\pi r^2 n v_c}
\end{equation}
where $\lambda$ is the mean free path, $r$ is the radius of the clouds, $n$ is the number density of clouds and $v_c$ the cloud-cloud velocity dispersion. The number density of clouds varies with Galactic radius, and scale height above the disc. For simplicity, here we study the disc within a radius of 5 kpc and take a scale height of 50 pc. This gives a number density of $8.8 \times 10^{-8}$ clouds per pc$^3$. Taking a typical cloud radius of 50 pc, and $v_c=4$ km $^{-1}$, we obtain $t_{coll} \sim 350$ Myr. This is similar to the estimate in Blitz \& Shu 1980, though the latter includes a gravitational term to increase the cross section. However if we only select the clouds in the spiral arms, the number density of clouds increases dramatically. By taking the arm width to be 150 pc (a high estimate), and using the pitch angle, the volume occupied by the arms is less than 5 \% of the overall disc. As the majority of clouds lie in the spiral arms, we obtain $t_{coll} \lesssim 20$ Myr (and $\lambda \lesssim 70$ pc) if we only include the spiral arms, which is in much better agreement with the results we find directly from the hydrodynamic calculations. Again, for the simulation without an imposed spiral potential, we would expect a higher rate of collisions theoretically if we reduced the volume to reflect the clustered nature of the clouds, compared to assuming a uniform distribution.

We do not consider the frequency of collisions of non-spherical clouds here, however collisions of non-spherical molecules have been studied by \citet{Gop2011}. They find an increase in the collision frequency by a factor of up to 5 for highly elongated molecules.

\section{Synthetic observations of mergers}
In this section we utilise the inclusion of H$_2$ and CO in our models to search for and show cloud mergers using a molecular rather than total density threshold. We take a molecular density threshold of 10 cm$^{-3}$ (see Table~1) which is quite low, but the amount of molecular gas at densities much higher than this is limited by our inclusion of stellar feedback, and gas at total densities $<100$ cm$^{-3}$ is not fully molecular. Thus our molecular gas criteria is similar to the criteria for total gas, because gas is often still not fully molecular (note also that here we primarily consider CO velocity information rather than intensities).   
We checked the frequency of mergers, and the fractions $f_0, f_1, f_2$ as for the earlier parts of the paper, and found similar results compared to our fiducial analysis with a $\rho_{min}$ for the total gas of 50 cm$^{-3}$.  This is not so surprising, as the fraction  of gas in clouds, and surface densities of the clouds using these different criteria are very similar. We also compared the example massive cloud mergers as found in Table~2. We tend not to find exactly the same mergers (though two were clearly identifiable as the same mergers as in Table~2) indicating that the details of interactions themselves may be sensitive to whether the total or molecular density is used. However we found a similar number of interactions of massive clouds (and interactions in total) using the molecular density criteria, and again most of the interactions involved one very filamentary cloud, and another joining on the end.

\begin{figure}
\centerline{\includegraphics[scale=0.45]{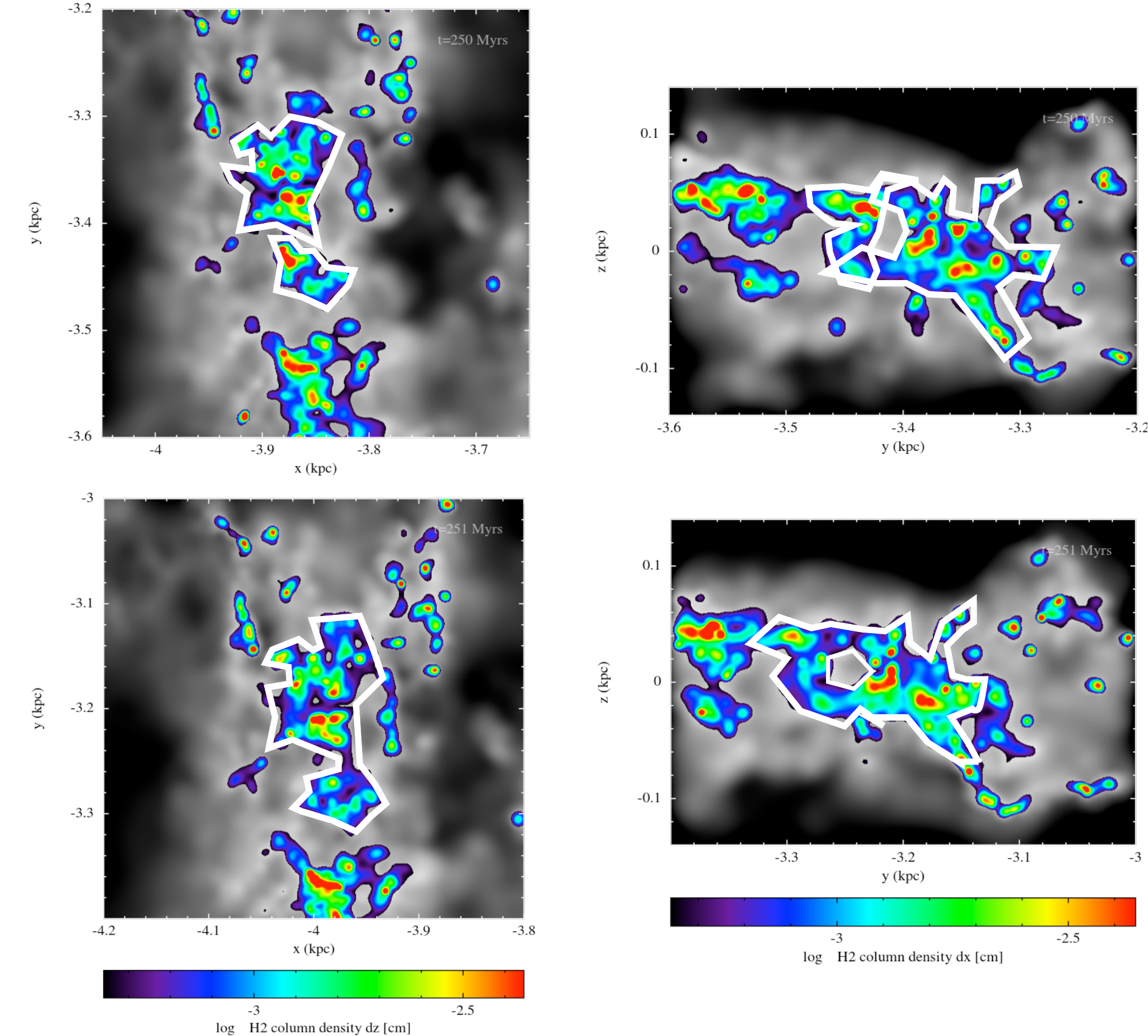}}
\caption{The merger of two massive clouds is shown, at times of 250 Myr (top) and 251 Myr (lower) and in the $xy$ plane (left) and the $yz$ plane. The colour density scale shows the H$_2$ column density, which is over plotted on the total column density in black and white. White contours indicate the cloud boundaries. In this case, the clouds were selected based on H$_2$ densities.}
\label{fig:h2}
\end{figure}

We show in Figure~\ref{fig:h2} the merger shown in Figure~\ref{fig:collision2} , but with clouds found from the H$_2$ density, rather than the total density. The merger looks fairly similar to the case with the total density, but there are subtle differences where the total density is high, but not the H$_2$, and vice versa. 

\begin{figure}
\centerline{\includegraphics[scale=0.33]{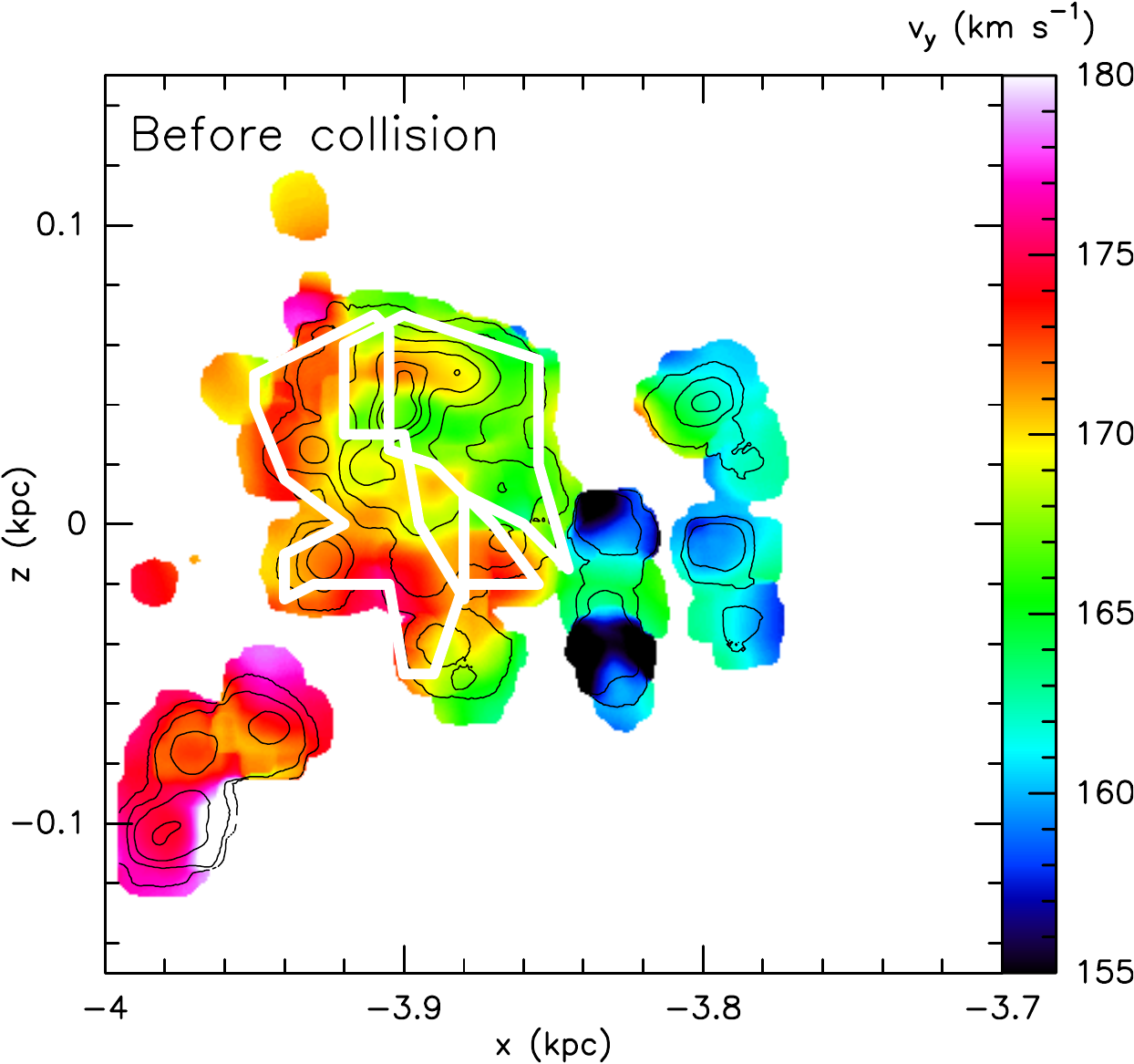}
\includegraphics[scale=0.33]{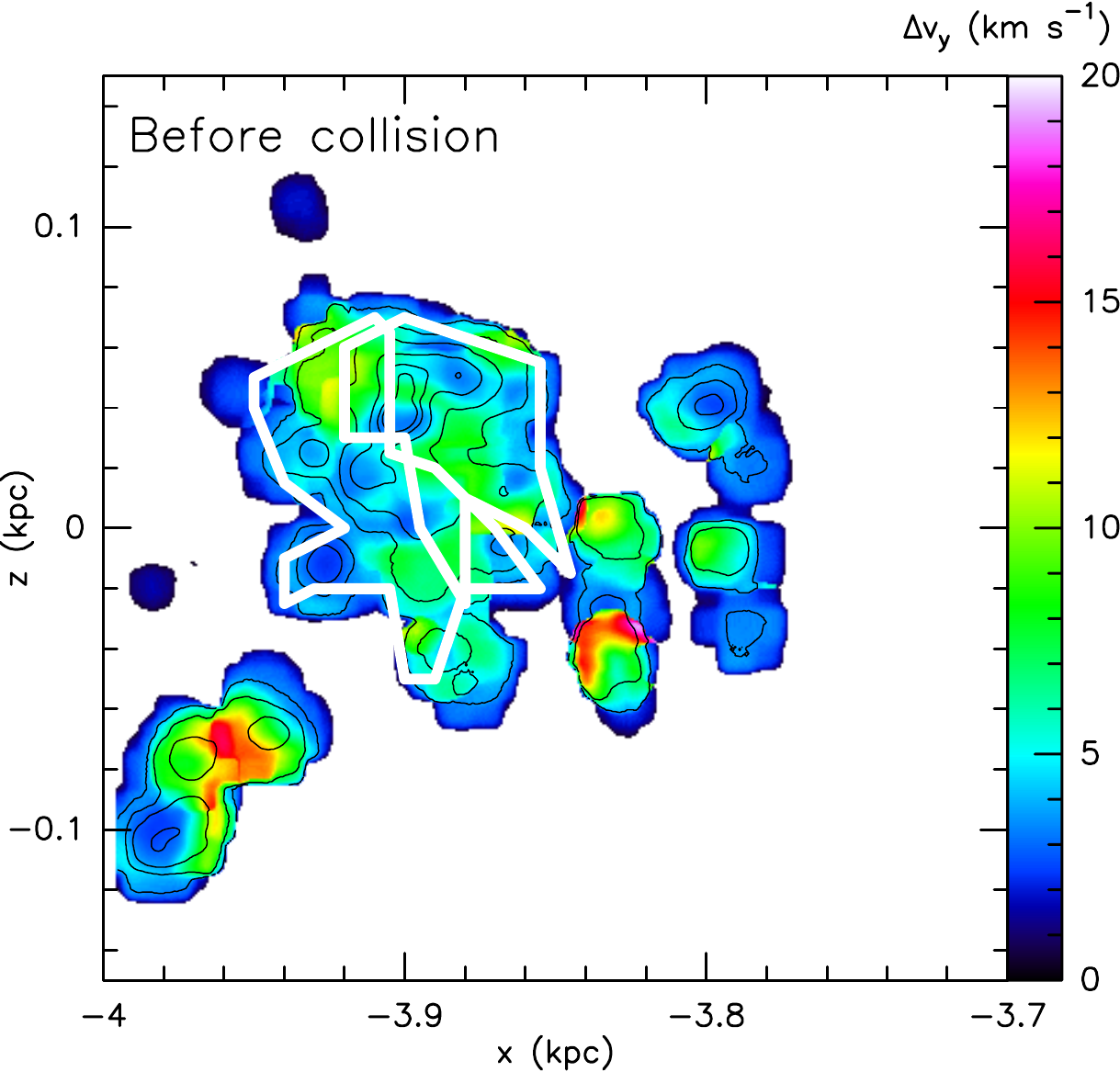}}
\centerline{\includegraphics[scale=0.33]{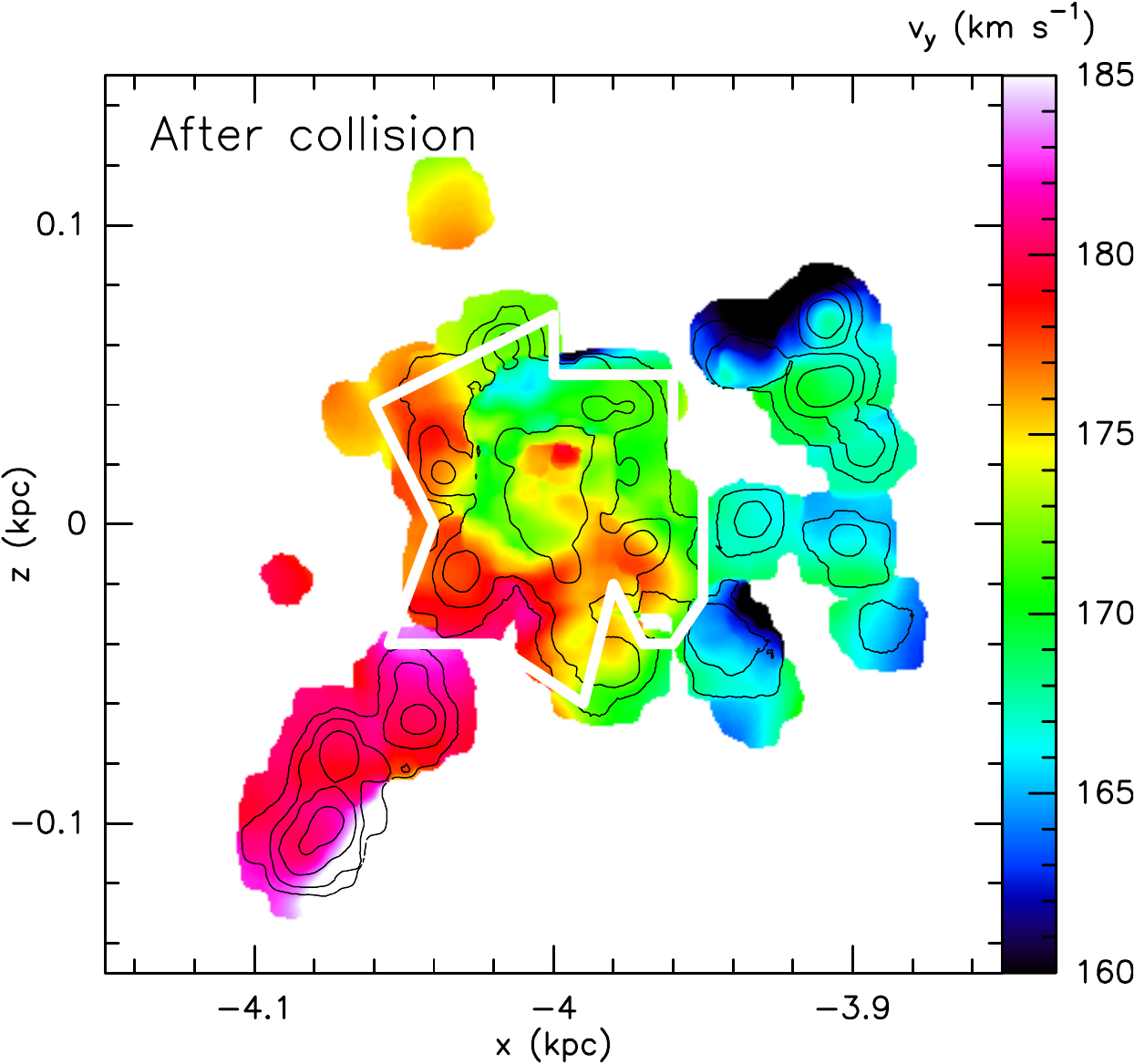}
\includegraphics[scale=0.33]{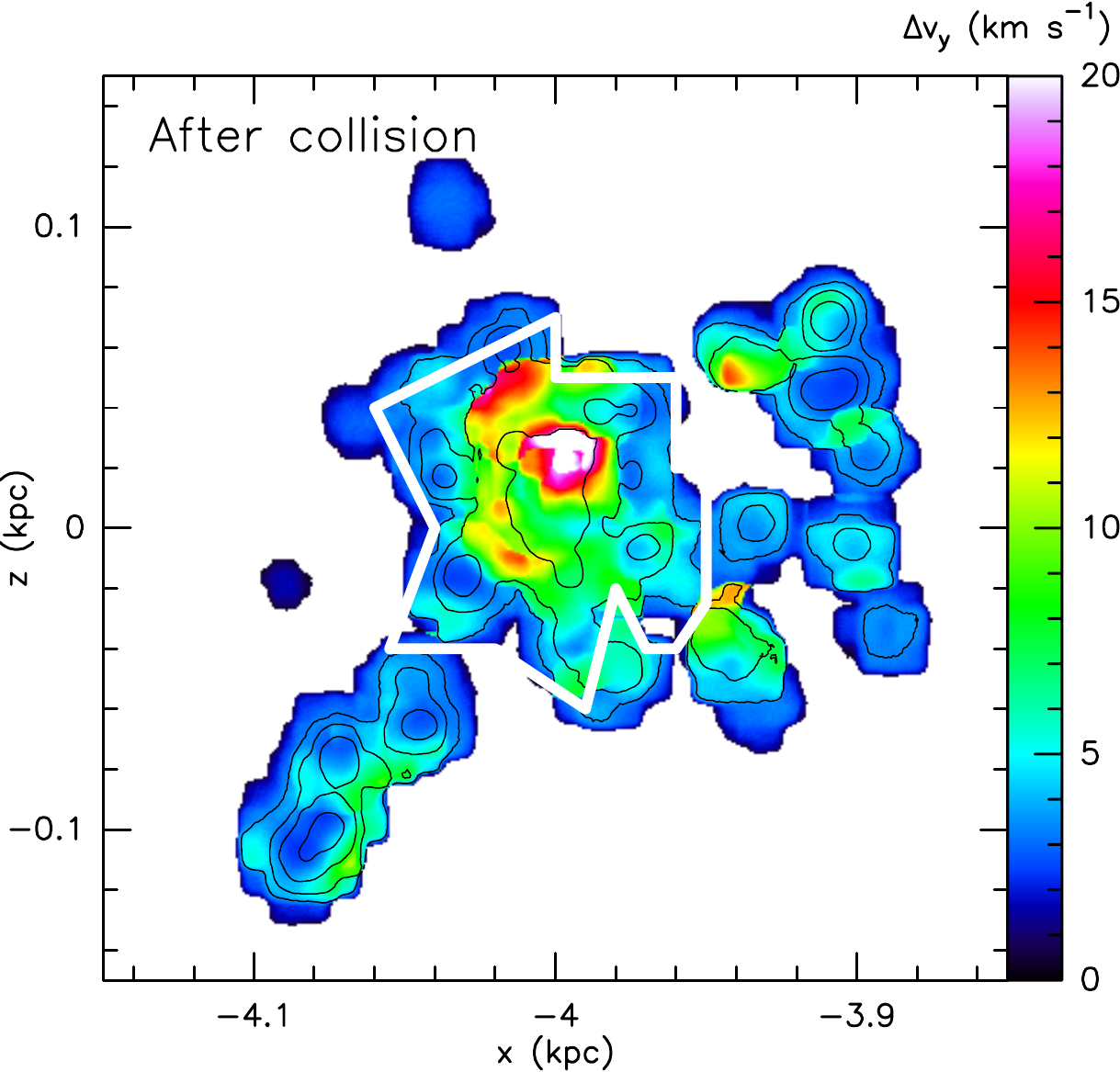}}
\caption{The merger of two massive clouds is shown, at times of 250 Myr (top) and 251 Myr (lower) in CO(1-0) integrated intensities (black contours). The colour scale shows $v_y$ (left) and the dispersion, $\Delta v_y$ (right). A velocity gradient is clear where the clouds are colliding. There is also an increase in the velocity dispersion after the clouds have collided. White contours indicate the approximate boundaries of the clouds (from the particles).}
\label{fig:co}
\end{figure}

To see how this merger appears in the CO (1-0) transition, we post-processed the results from the simulation using the TORUS radiative transfer code (\citealt{Harries2000, Acreman2012}, Duarte-Cabral et al 2014 submitted). In Figure~\ref{fig:co} , we show the CO emission of these clouds in $xz$ space (not shown in Figure~\ref{fig:h2}) so that the line of sight component of the velocity corresponds to the velocity along the $y$ axis, which is the direction in which the two clouds are colliding (see Figure~\ref{fig:h2}). In the $xz$ plane, the clouds are not obviously physically separated, but one is clearly at lower values of $x$ compared to the other. 

The plots showing the velocity field (left) clearly show two different velocities for the two clouds, which are kept even after the collision occurs, with a relatively strong velocity discontinuity where the collision occurs ($\sim$10 km s$^{-1}$). These velocity field maps were created by calculating the first moment of the CO emission, and therefore, they do not recover all the complexity of the velocity structure. One way to assess the complexity of the velocity structure, is by assessing the line of sight velocity dispersion ($\Delta_{v_y}$)), calculated as the second moment of the CO emission (right hand plots of Figure 14). These panels show a region of high, albeit localised velocity dispersion which is present after the collision but not before, reaching as high as 20km s$^{-1}$ (more than doubling the original $\Delta_{v_y}$ of the clouds). We also built a couple of position-velocity cuts (not shown) running across this high velocity dispersion region (at z$\sim$0.02 kpc), which confirm that the velocity structure before the collision is quite simple, with smooth and clearly separated velocities, while the velocity structure after the collision shows two or even three velocity components, where the velocity dispersion is higher.

We note, however, that these signatures are only distinctive when we can observe the collision along the collision axis. For instance, our synthetic observations of the $yz$ plane (also not shown) do not show any significant change of velocity dispersions (along the $x$ axis) after the collision, as there is minimal mixing along that axis. As collisions in the Galaxy are rarely oriented along the observer's line of sight, the velocity discontinuity and velocity dispersion that we estimate for this particular collision should be taken as indicative (upper) limits of what would be observed, as they are measured along the collision axis. 

Overall, we find that such collisions do reproduce the signatures that observers attribute to cloud-cloud collisions, such as velocity shears with several overlapping velocity components (as seen in position-velocity diagrams, e.g. \citealt{Fukui2014,Duarte2010}), and increased line widths (by more than a factor two). Although most observations of cloud collisions in the Galaxy are of lower-mass molecular clouds and lower collision velocities (of only a couple of km s$^{-1}$) compared to the GMCs collisions we study here, there is evidence of similar GMCs collisions in the Galaxy, as that reported by \citet{Fukui2014}, with a velocity discontinuity of $\sim$ 15km s$^{-1}$ and $\Delta_v$ of $\gtrsim$10km s$^{-1}$. 

In conclusion, although the morphology/density structure of GMCs in the simulations do not suffer a great change due to a collision event, collisions do have an impact on the global dynamics of the clouds, which could potentially have an impact on the star formation taking place (that we do not resolve with the current resolution). We again note, as for Section 4, that the number of highly resolved clouds where we can construct this kind of analysis is limited by the resolution of the simulation.

\section{Conclusions}
We have studied the evolution of GMCs over short time periods (0.1-5 Myr) using galactic simulations of the ISM. Cloud evolution can be divided into a complete set of 5 categories: No change, Create, Merge, Split, Destroy. Up to timescales of 5 Myr, the most frequent evolution of a cloud is `No change'. This timescale also corresponds to the timescale over which the interaction of clouds with intercloud material  starts to become substantial, and needs to be taken into account. A timescale of 1 Myr is appropriate for studying merges and splits, as these processes reduce to 2 body problems (longer timeframes allow multiple mergers and interactions). The frequency of mergers of clouds $>10^4$ M$_{\odot}$ is about one per 8-10 Myr, or 1 per 15th of an orbit. This results in typically one merger per cloud lifetime, possibly none for the shortest--lived clouds or 2 or 3 for longer--lived (often more massive) clouds. In the absence of spiral arms, this reduces to one merger per $\sim$ 28 Myr (1 per 1/5th of an orbit, in good agreement with \citealt{Tasker2009}). Both are more frequent than previous analytic estimates, even in the case without spiral arms, as clouds are unevenly distributed throughout the galaxy.

Although we find that mergers or collisions are relatively frequent in the simulations, and at any point in time we can find a number of examples as illustrated by Section~4, they do not appear to have much impact. The reason for this is partly due to cloud orientations, as clouds are often elongated, and aligned with spiral arms. So, counter-intuitively, mergers are often prone to occur along the minor axes of the clouds, rather than the major axes (where we denote the minor axes in relation to the cross section of the clouds, as in Figure~9a), thus only effecting smaller areas of the clouds. Furthermore the velocity dispersions between clouds are not that high. These factors mean that cloud mergers or collisions appear more often simply as one cloud `nudging' another, with little or no change in the global density structure. Consequently although the clouds may `merge' and the collective mass increases (and the global velocity field changes), there appears to be little mixing of the gas in the two clouds. Rather each cloud retains its own characteristics. An expression such as `collision', which implies some significant change to one or both clouds' structures, may therefore be inappropriate to describe interactions of clouds, whilst `mergers' or simply `interactions' may be preferable. The cloud--cloud interactions also do not resemble particularly the colliding flow scenario of cloud evolution and star formation. Our picture is however in agreement with previous studies \citep{Elmegreen1986,Dobbs2011new} which suppose that spiral arms make little difference to the star formation rate, as the increase of cloud--cloud interactions we find in spiral arms has little impact on the ISM, except to group dense gas into larger structures.

Our simulations do not show any evidence that collisions of massive clouds could be responsible for massive clusters. Although the frequency of massive cloud interactions is not prohibitive (the number of mergers roughly corresponds to the numbers of very massive clusters), the interactions appear not be violent or quick enough, and often involve involve small parts of the clouds.
One difference between the observations of massive cloud collisions, and our simulations, is that the relative velocities of the observed collisions (e.g. 20 km s$^{-1}$, \citealt{Fukui2014}) lie at the extreme end of the examples we examine, and would expect from the cloud--cloud velocity dispersions in our and nearby galaxies. 

Lastly we note some caveats to our results though that would ideally be considered in future work. One caveat is that we do not have the resolution to study cloud--cloud interactions in detail. In particular we are limited by the density threshold for adding feedback, so may miss large increases of density  where the clouds collide. This could allow massive clusters of short age spreads, but our simulations suggest this would only occur at the the interface of considerably more extended clouds. We also have only a simple feedback scheme. However most of our results, e.g. the relative frequency of cloud mergers in different cases, the tendency of clouds to be elongated and aligned when colliding, and the cloud--cloud velocity dispersions, are not likely to depend strongly on feedback. Furthermore \citet{Tasker2009} achieve similar collision frequencies with no stellar feedback. Our timescale of 5~Myr, denoting when cloud and intercloud material start to substantially may be more subject to the details of feedback.
A third caveat is that we have not considered more extreme environments (e.g. galaxy mergers, high redshift galaxies, the Galactic Centre), which could potentially be more conducive to more violent cloud--cloud interactions. We note that of the observational and numerical studies related to massive cloud-cloud collisions, a number concern Galactic Centre clouds (e.g. \citealt{Stolte2008,Hobbs2009,Johnston2014}).

\section{Acknowledgments}
The calculations for this paper were performed on the DiRAC machine `Complexity', and the supercomputer at Exeter, which is jointly funded by STFC, the Large Facilities Capital Fund of BIS, and the University of Exeter. Figures~2, 11, 12 and 14 were produced using \textsc{splash} \citep{splash2007}.
We thank the referee, Robi Banerjee, for a helpful report which improved our explanations in some parts of the paper. CLD and ADC acknowledge funding from the European Research Council for the FP7 ERC starting grant project LOCALSTAR. CLD thanks Thomas Henning, Annie Hughes, Steve Longmore and Jin Koda for useful comments and discussions.
\bibliographystyle{mn2e}
\bibliography{Dobbs}

\begin{thebibliography}{}

\bibitem[\protect\citeauthoryear{{Acreman}, {Dobbs}, {Brunt} \&
  {Douglas}}{{Acreman} et~al.}{2012}]{Acreman2012}
{Acreman} D.~M.,  {Dobbs} C.~L.,  {Brunt} C.~M.,    {Douglas} K.~A.,  2012,
  \mnras, 422, 241

\bibitem[\protect\citeauthoryear{{Bate}, {Bonnell} \& {Price}}{{Bate}
  et~al.}{1995}]{Bate1995}
{Bate} M.~R.,  {Bonnell} I.~A.,    {Price} N.~M.,  1995, \mnras, 277, 362

\bibitem[\protect\citeauthoryear{{Bekki}, {Beasley}, {Forbes} \&
  {Couch}}{{Bekki} et~al.}{2004}]{Bekki2004}
{Bekki} K.,  {Beasley} M.~A.,  {Forbes} D.~A.,    {Couch} W.~J.,  2004, \apj,
  602, 730

\bibitem[\protect\citeauthoryear{{Benz}, {Cameron}, {Press} \& {Bowers}}{{Benz}
  et~al.}{1990}]{Benz1990}
{Benz} W.,  {Cameron} A.~G.~W.,  {Press} W.~H.,    {Bowers} R.~L.,  1990, \apj,
  348, 647

\bibitem[\protect\citeauthoryear{{Blitz} \& {Shu}}{{Blitz} \&
  {Shu}}{1980}]{Blitz1980}
{Blitz} L.,  {Shu} F.~H.,  1980, \apj, 238, 148

\bibitem[\protect\citeauthoryear{{Bonnell}, {Dobbs}, {Robitaille} \&
  {Pringle}}{{Bonnell} et~al.}{2006}]{Bonnell2006}
{Bonnell} I.~A.,  {Dobbs} C.~L.,  {Robitaille} T.~R.,    {Pringle} J.~E.,
  2006, \mnras, 365, 37

\bibitem[\protect\citeauthoryear{{Casoli} \& {Combes}}{{Casoli} \&
  {Combes}}{1982}]{Casoli1982}
{Casoli} F.,  {Combes} F.,  1982, \aap, 110, 287

\bibitem[\protect\citeauthoryear{{Colombo}, {Hughes}, {Schinnerer}, {Meidt},
  {Leroy}, {Pety}, {Dobbs}, {Garc{\'{\i}}a-Burillo}, {Dumas}, {Thompson},
  {Schuster} \& {Kramer}}{{Colombo} et~al.}{2014}]{Colombo2014}
{Colombo} D.,  {Hughes} A.,  {Schinnerer} E.,  {Meidt} S.~E.,  {Leroy} A.~K.,
  {Pety} J.,  {Dobbs} C.~L.,  {Garc{\'{\i}}a-Burillo} S.,  {Dumas} G.,
  {Thompson} T.~A.,  {Schuster} K.~F.,    {Kramer} C.,  2014, \apj, 784, 3

\bibitem[\protect\citeauthoryear{{Cox} \& {G{\' o}mez}}{{Cox} \& {G{\'
  o}mez}}{2002}]{Cox2002}
{Cox} D.~P.,  {G{\' o}mez} G.~C.,  2002, \apjs, 142, 261

\bibitem[\protect\citeauthoryear{{Dobbs}}{{Dobbs}}{2008}]{Dobbs2008}
{Dobbs} C.~L.,  2008, \mnras, 391, 844

\bibitem[\protect\citeauthoryear{{Dobbs} \& {Bonnell}}{{Dobbs} \&
  {Bonnell}}{2007}]{Dobbs2007a}
{Dobbs} C.~L.,  {Bonnell} I.~A.,  2007, \mnras, 374, 1115

\bibitem[\protect\citeauthoryear{{Dobbs}, {Burkert} \& {Pringle}}{{Dobbs}
  et~al.}{2011}]{Dobbs2011new}
{Dobbs} C.~L.,  {Burkert} A.,    {Pringle} J.~E.,  2011, \mnras, 417, 1318

\bibitem[\protect\citeauthoryear{{Dobbs} \& {Pringle}}{{Dobbs} \&
  {Pringle}}{2013}]{Dobbs2013}
{Dobbs} C.~L.,  {Pringle} J.~E.,  2013, \mnras, 432, 653

\bibitem[\protect\citeauthoryear{{Dobbs}, {Pringle} \& {Naylor}}{{Dobbs}
  et~al.}{2014}]{Dobbs2014}
{Dobbs} C.~L.,  {Pringle} J.~E.,    {Naylor} T.,  2014, \mnras, 437, L31

\bibitem[\protect\citeauthoryear{{Duarte-Cabral}, {Dobbs}, {Peretto} \&
  {Fuller}}{{Duarte-Cabral} et~al.}{2011}]{Duarte2011}
{Duarte-Cabral} A.,  {Dobbs} C.~L.,  {Peretto} N.,    {Fuller} G.~A.,  2011,
  \aap, 528, A50

\bibitem[\protect\citeauthoryear{{Duarte-Cabral}, {Fuller}, {Peretto},
  {Hatchell}, {Ladd}, {Buckle}, {Richer} \& {Graves}}{{Duarte-Cabral}
  et~al.}{2010}]{Duarte2010}
{Duarte-Cabral} A.,  {Fuller} G.~A.,  {Peretto} N.,  {Hatchell} J.,  {Ladd}
  E.~F.,  {Buckle} J.,  {Richer} J.,    {Graves} S.~F.,  2010, \aap, 519, A27

\bibitem[\protect\citeauthoryear{{Elmegreen} \& {Elmegreen}}{{Elmegreen} \&
  {Elmegreen}}{1986}]{Elmegreen1986}
{Elmegreen} B.~G.,  {Elmegreen} D.~M.,  1986, \apj, 311, 554

\bibitem[\protect\citeauthoryear{{Field} \& {Saslaw}}{{Field} \&
  {Saslaw}}{1965}]{Field1965}
{Field} G.~B.,  {Saslaw} W.~C.,  1965, \apj, 142, 568

\bibitem[\protect\citeauthoryear{{Fujimoto}, {Tasker}, {Wakayama} \&
  {Habe}}{{Fujimoto} et~al.}{2014}]{Fujimoto2014}
{Fujimoto} Y.,  {Tasker} E.~J.,  {Wakayama} M.,    {Habe} A.,  2014, \mnras,
  439, 936

\bibitem[\protect\citeauthoryear{{Fukui, et al.}}{{Fukui, et
  al.}}{2014}]{Fukui2014}
{Fukui, et al.} 2014, \apj, 780, 36

\bibitem[\protect\citeauthoryear{{Furukawa}, {Dawson}, {Ohama}, {Kawamura},
  {Mizuno}, {Onishi} \& {Fukui}}{{Furukawa} et~al.}{2009}]{Furukawa2009}
{Furukawa} N.,  {Dawson} J.~R.,  {Ohama} A.,  {Kawamura} A.,  {Mizuno} N.,
  {Onishi} T.,    {Fukui} Y.,  2009, \apjl, 696, L115

\bibitem[\protect\citeauthoryear{{Galv{\'a}n-Madrid}, {Zhang}, {Keto}, {Ho},
  {Zapata}, {Rodr{\'{\i}}guez}, {Pineda} \&
  {V{\'a}zquez-Semadeni}}{{Galv{\'a}n-Madrid} et~al.}{2010}]{Galvan2010}
{Galv{\'a}n-Madrid} R.,  {Zhang} Q.,  {Keto} E.,  {Ho} P.~T.~P.,  {Zapata}
  L.~A.,  {Rodr{\'{\i}}guez} L.~F.,  {Pineda} J.~E.,    {V{\'a}zquez-Semadeni}
  E.,  2010, \apj, 725, 17

\bibitem[\protect\citeauthoryear{{Glover} \& {Mac Low}}{{Glover} \& {Mac
  Low}}{2007}]{Glover2007}
{Glover} S.~C.~O.,  {Mac Low} M.-M.,  2007, \apjs, 169, 239

\bibitem[\protect\citeauthoryear{{Gopalakrishnan}, {Thajudeen} \&
  {Hogan}}{{Gopalakrishnan} et~al.}{2011}]{Gop2011}
{Gopalakrishnan} R.,  {Thajudeen} T.,    {Hogan} C.~J.,  2011, \jcp, 135,
  054302

\bibitem[\protect\citeauthoryear{{Harries}}{{Harries}}{2000}]{Harries2000}
{Harries} T.~J.,  2000, \mnras, 315, 722

\bibitem[\protect\citeauthoryear{{Hausman}}{{Hausman}}{1981}]{Hausman1981}
{Hausman} M.~A.,  1981, \apj, 245, 72

\bibitem[\protect\citeauthoryear{{Heyer}, {Krawczyk}, {Duval} \&
  {Jackson}}{{Heyer} et~al.}{2009}]{Heyer2009}
{Heyer} M.,  {Krawczyk} C.,  {Duval} J.,    {Jackson} J.~M.,  2009, \apj, 699,
  1092

\bibitem[\protect\citeauthoryear{{Higuchi}, {Kurono}, {Saito} \&
  {Kawabe}}{{Higuchi} et~al.}{2010}]{Higuchi2010}
{Higuchi} A.~E.,  {Kurono} Y.,  {Saito} M.,    {Kawabe} R.,  2010, \apj, 719,
  1813

\bibitem[\protect\citeauthoryear{{Hobbs} \& {Nayakshin}}{{Hobbs} \&
  {Nayakshin}}{2009}]{Hobbs2009}
{Hobbs} A.,  {Nayakshin} S.,  2009, \mnras, 394, 191

\bibitem[\protect\citeauthoryear{{Inoue} \& {Fukui}}{{Inoue} \&
  {Fukui}}{2013}]{Inoue2013}
{Inoue} T.,  {Fukui} Y.,  2013, \apjl, 774, L31

\bibitem[\protect\citeauthoryear{{Johnston}, {Beuther}, {Linz}, {Schmiedeke},
  {Ragan} \& {Henning}}{{Johnston} et~al.}{2014}]{Johnston2014}
{Johnston} K.~G.,  {Beuther} H.,  {Linz} H.,  {Schmiedeke} A.,  {Ragan} S.~E.,
    {Henning} T.,  2014, \aap, 568, A56

\bibitem[\protect\citeauthoryear{{Kennicutt}}{{Kennicutt}}{1998}]{Kennicutt1998}
{Kennicutt} R.~C.,  1998, \araa, 36, 189

\bibitem[\protect\citeauthoryear{{Kitsionas} \& {Whitworth}}{{Kitsionas} \&
  {Whitworth}}{2007}]{Kitsionas2007}
{Kitsionas} S.,  {Whitworth} A.~P.,  2007, \mnras, 378, 507

\bibitem[\protect\citeauthoryear{{Kwan} \& {Valdes}}{{Kwan} \&
  {Valdes}}{1983}]{Kwan1983}
{Kwan} J.,  {Valdes} F.,  1983, \apj, 271, 604

\bibitem[\protect\citeauthoryear{{Kwan} \& {Valdes}}{{Kwan} \&
  {Valdes}}{1987}]{Kwan1987}
{Kwan} J.,  {Valdes} F.,  1987, \apj, 315, 92

\bibitem[\protect\citeauthoryear{{Lattanzio}, {Monaghan}, {Pongracic} \&
  {Schwarz}}{{Lattanzio} et~al.}{1985}]{Lattanzio1985}
{Lattanzio} J.~C.,  {Monaghan} J.~J.,  {Pongracic} H.,    {Schwarz} M.~P.,
  1985, \mnras, 215, 125

\bibitem[\protect\citeauthoryear{{Longmore}, {Kruijssen}, {Bastian}, {Bally},
  {Rathborne}, {Testi}, {Stolte}, {Dale}, {Bressert} \& {Alves}}{{Longmore}
  et~al.}{2014}]{Longmore2014}
{Longmore} S.~N.,  {Kruijssen} J.~M.~D.,  {Bastian} N.,  {Bally} J.,
  {Rathborne} J.,  {Testi} L.,  {Stolte} A.,  {Dale} J.,  {Bressert} E.,
  {Alves} J.,  2014, ArXiv 1401 4175

\bibitem[\protect\citeauthoryear{{McLeod}, {Palou{\v s}} \&
  {Whitworth}}{{McLeod} et~al.}{2011}]{McLeod2011}
{McLeod} A.,  {Palou{\v s}} J.,    {Whitworth} A.,  2011, in {Alves} J.,
  {Elmegreen} B.~G.,  {Girart} J.~M.,   {Trimble} V.,  eds, Computational Star
  Formation Vol.~270 of IAU Symposium, {Collisions of supersonic clouds}.
pp 355--358

\bibitem[\protect\citeauthoryear{{Nakamura}, {Miura}, {Kitamura}, {Shimajiri},
  {Kawabe}, {Akashi}, {Ikeda}, {Tsukagoshi}, {Momose}, {Nishi} \&
  {Li}}{{Nakamura} et~al.}{2012}]{Nakamura2012}
{Nakamura} F.,  {Miura} T.,  {Kitamura} Y.,  {Shimajiri} Y.,  {Kawabe} R.,
  {Akashi} T.,  {Ikeda} N.,  {Tsukagoshi} T.,  {Momose} M.,  {Nishi} R.,
  {Li} Z.-Y.,  2012, \apj, 746, 25

\bibitem[\protect\citeauthoryear{{Norman} \& {Silk}}{{Norman} \&
  {Silk}}{1980}]{Norman1980}
{Norman} C.,  {Silk} J.,  1980, \apj, 238, 158

\bibitem[\protect\citeauthoryear{{Pettitt}, {Dobbs}, {Acreman} \&
  {Price}}{{Pettitt} et~al.}{2014}]{Pettitt2014}
{Pettitt} A.~R.,  {Dobbs} C.~L.,  {Acreman} D.~M.,    {Price} D.~J.,  2014,
  \mnras, 444, 919

\bibitem[\protect\citeauthoryear{{Price}}{{Price}}{2007}]{splash2007}
{Price} D.~J.,  2007, Publications of the Astronomical Society of Australia,
  24, 159

\bibitem[\protect\citeauthoryear{{Price} \& {Monaghan}}{{Price} \&
  {Monaghan}}{2007}]{PM2007}
{Price} D.~J.,  {Monaghan} J.~J.,  2007, \mnras, 374, 1347

\bibitem[\protect\citeauthoryear{{Roberts} \& {Stewart}}{{Roberts} \&
  {Stewart}}{1987}]{Roberts1987}
{Roberts} W.~W.,  {Stewart} G.~R.,  1987, \apj, 314, 10

\bibitem[\protect\citeauthoryear{{Scoville}, {Solomon} \& {Sanders}}{{Scoville}
  et~al.}{1979}]{Scoville1979}
{Scoville} N.~Z.,  {Solomon} P.~M.,    {Sanders} D.~B.,  1979, in {Burton}
  W.~B.,  ed., The Large-Scale Characteristics of the Galaxy Vol.~84 of IAU
  Symposium, {CO observations of spiral structure and the lifetime of giant
  molecular clouds}.
pp 277--282

\bibitem[\protect\citeauthoryear{{Silk}}{{Silk}}{1997}]{Silk1997}
{Silk} J.,  1997, \apj, 481, 703

\bibitem[\protect\citeauthoryear{{Stark} \& {Lee}}{{Stark} \&
  {Lee}}{2005}]{Stark2005}
{Stark} A.~A.,  {Lee} Y.,  2005, \apjl, 619, L159

\bibitem[\protect\citeauthoryear{{Stolte}, {Ghez}, {Morris}, {Lu}, {Brandner}
  \& {Matthews}}{{Stolte} et~al.}{2008}]{Stolte2008}
{Stolte} A.,  {Ghez} A.~M.,  {Morris} M.,  {Lu} J.~R.,  {Brandner} W.,
  {Matthews} K.,  2008, \apj, 675, 1278

\bibitem[\protect\citeauthoryear{{Taff} \& {Savedoff}}{{Taff} \&
  {Savedoff}}{1973}]{Taff1973}
{Taff} L.~G.,  {Savedoff} M.~P.,  1973, \mnras, 164, 357

\bibitem[\protect\citeauthoryear{{Tan}}{{Tan}}{2000}]{Tan2000}
{Tan} J.~C.,  2000, \apj, 536, 173

\bibitem[\protect\citeauthoryear{{Tasker} \& {Tan}}{{Tasker} \&
  {Tan}}{2009}]{Tasker2009}
{Tasker} E.~J.,  {Tan} J.~C.,  2009, \apj, 700, 358

\bibitem[\protect\citeauthoryear{{Tomisaka}}{{Tomisaka}}{1984}]{Tomisaka1984}
{Tomisaka} K.,  1984, \pasj, 36, 457

\bibitem[\protect\citeauthoryear{{Tomisaka}}{{Tomisaka}}{1986}]{Tomisaka1986}
{Tomisaka} K.,  1986, \pasj, 38, 95

\bibitem[\protect\citeauthoryear{{Torii}, {Enokiya}, {Sano}, {Yoshiike},
  {Hanaoka}, {Ohama}, {Furukawa}, {Dawson}, {Moribe}, {Oishi}, {Nakashima},
  {Okuda}, {Yamamoto}, {Kawamura}, {Mizuno}, {Maezawa}, {Onishi}, {Mizuno} \&
  {Fukui}}{{Torii} et~al.}{2011}]{Torii2011}
{Torii} K.,  {Enokiya} R.,  {Sano} H.,  {Yoshiike} S.,  {Hanaoka} N.,  {Ohama}
  A.,  {Furukawa} N.,  {Dawson} J.~R.,  {Moribe} N.,  {Oishi} K.,  {Nakashima}
  Y.,  {Okuda} T.,  {Yamamoto} H.,  {Kawamura} A.,  {Mizuno} N.,  {Maezawa} H.,
   {Onishi} T.,  {Mizuno} A.,    {Fukui} Y.,  2011, \apj, 738, 46

\bibitem[\protect\citeauthoryear{{V{\'a}zquez-Semadeni}, {G{\'o}mez},
  {Jappsen}, {Ballesteros-Paredes}, {Gonz{\'a}lez} \&
  {Klessen}}{{V{\'a}zquez-Semadeni} et~al.}{2007}]{Vaz2007}
{V{\'a}zquez-Semadeni} E.,  {G{\'o}mez} G.~C.,  {Jappsen} A.~K.,
  {Ballesteros-Paredes} J.,  {Gonz{\'a}lez} R.~F.,    {Klessen} R.~S.,  2007,
  \apj, 657, 870

\bibitem[\protect\citeauthoryear{{Wilson et al.}}{{Wilson et
  al.}}{2011}]{Wilson2011}
{Wilson et al.} 2011, \mnras, 410, 1409

\bibitem[\protect\citeauthoryear{{Wyse}}{{Wyse}}{1986}]{Wyse1986}
{Wyse} R.~F.~G.,  1986, \apjl, 311, L41

\bibitem[\protect\citeauthoryear{{Wyse} \& {Silk}}{{Wyse} \&
  {Silk}}{1989}]{Wyse1989}
{Wyse} R.~F.~G.,  {Silk} J.,  1989, \apj, 339, 700

\end{thebibliography}

\bsp
\label{lastpage}
\end{document}